\documentclass[aps,prd,twocolumn,nofootinbib]{revtex4}
\pdfoutput=1
\usepackage{epsfig}
\usepackage{amsmath}
\usepackage{graphicx}
\usepackage{booktabs}
\usepackage{multirow}
\usepackage{rotating}
\newcommand{\beq}{\begin{equation}}
\newcommand{\eeq}{\end{equation}}
\newcommand{\bqa}{\begin{eqnarray}}
\newcommand{\eqa}{\end{eqnarray}}

\newcommand{\bold}{\textbf}

\begin{document}

\title{Analyzing doubly heavy tetra- and penta-quark states by variational method}
\author{ Ruilin Zhu$^{1}$,  Xuejie Liu$^{1}$, Hongxia Huang$^{1}$,
Cong-Feng Qiao$^{2,3}$~\footnote{Corresponding author:qiaocf@ucas.edu.cn}\vspace{0.2cm}}
\affiliation{$^1$Department of Physics and Institute of Theoretical Physics,
Nanjing Normal University, Nanjing, Jiangsu 210023, China\\
$^2$School of Physics, University of Chinese Academy of Sciences,
YuQuan Road 19A, Beijing 100049, China\\
$^3$CAS Center for Excellence in Particle Physics, Beijing 100049, China}

\begin{abstract}
Motivated by the very recent observations of hidden charm pentaquarks $P_c(4312)^+$, $P_c(4440)^+$ and $P_c(4457)^+$ of the LHCb Collaboration, we systematically study the spectra of the doubly-heavy (with or without charm/bottom numbers) pentaquarks and tetraquarks in non-relativistic constituent quark model. The model independent variational method is employed to solve the Schr\"odinger equation, where the test functions adopted are symmetric for the light quarks. In our study, the $P_c(4312)^+$ may be assigned as the ground state with spin-parity $\frac{1}{2}^-$ or $\frac{3}{2}^-$, while the $P_c(4440)^+$ and $P_c(4457)^+$ may be assigned as the excited states with $\frac{1}{2}^-$, which might all belong to the sextet with $s_{c\bar{c}}=1$ and $s_\ell=\frac{3}{2}$. It is notable that our working framework is quite similar to that of Hydrogen molecule, but with different potential structure. We also classify these pentaquarks and tetraquarks in light of the heavy quark symmetry and their decay properties are analyzed. Several promising channels for the observation of doubly-heavy pentaquarks and doubly-heavy tetraquarks in experiment are proposed.

\end{abstract}

\maketitle

\section{Introduction}

The discovery of exotic states greatly enriches the hadron family and our knowledge of the nature of QCD.
Up to date, more than thirty exotic states or candidates, denoted as $XYZ$ and $P_c$ states, have been observed in experiment.
To understand the properties of those exotic states and find more possible states are urgent tasks in hadron physics. However,
it seems the journey of exotic baryon study has just begun.

Very recently, the LHCb Collaboration announced the observations of hidden charm pentaquarks
$P_c(4312)^+$, $P_c(4440)^+$ and $P_c(4457)^+$ by the $P_c(X)^+\to J/\psi+p^+$ in
$\Lambda_b\to J/\psi+p^+ + K^- $ decays~\cite{LHCb2019,Aaij:2019vzc}. The data are consistent with
the 2015 results which led to two pentaquarks wide $P_c(4380)^+$ and narrow
$P_c(4450)^+$~\cite{Aaij:2015tga}. New data indicated that the previous $P_c(4450)^+$
bump actually belongs to two states $P_c(4440)^+$ and $P_c(4457)^+$, while $P_c(4312)^+$
is a novel candidate of hidden charm pentaquark. Their masses and decay widths go as \cite{Aaij:2019vzc}:
\begin{eqnarray*}
M_{P_c(4312)^+}&=&4311.9\pm0.7^{+6.8}_{-0.6}\ {\rm MeV},\\
\Gamma_{P_c(4312)^+}&=&9.8\pm2.7^{+3.7}_{-4.5}\ {\rm MeV};
\end{eqnarray*}
\begin{eqnarray*}
M_{P_c(4440)^+}&=&4440.3\pm1.3^{+4.1}_{-4.7}\ {\rm MeV},\\
\Gamma_{P_c(4440)^+}&=&20.6\pm4.9^{+8.7}_{-10.1}\ {\rm MeV};
\end{eqnarray*}
\begin{eqnarray*}
M_{P_c(4457)^+}&=&4457.3\pm0.6^{+4.1}_{-1.7}\ {\rm MeV},\\
\Gamma_{P_c(4457)^+}&=&6.4\pm2.0^{+5.7}_{-1.9}\ {\rm MeV}.
\end{eqnarray*}

There appear several theoretical interpretations for these three hidden charm pentaquarks in the literature \cite{Chen:2019asm,Chen:2019bip,Liu:2019tjn,Guo:2019fdo,He:2019ify,Liu:2019zoy,Xiao:2019mvs,Ali:2019npk,
Shimizu:2019ptd,Huang:2019jlf,Guo:2019kdc,Cao:2019kst,Weng:2019ynv,Mutuk:2019snd}. In Ref. \cite{Chen:2019asm}, the three narrow structures $P_c(4312)^+$, $P_c(4440)^+$ and $P_c(4457)^+$ were depicted as the molecular $\Sigma_c\bar{D}$ with (I=1/2, J=1/2), $\Sigma_c\bar{D}^*$ with (I=1/2, J=1/2) and  $\Sigma_c\bar{D}^*$ with (I=1/2, J=3/2), respectively. In Ref. \cite{Chen:2019bip}, $P_c(4457)^+$ was suggested to be the molecular $\Sigma_c^*\bar{D}^*$ with (I=1/2, J=5/2). Authors of Refs.~\cite{Liu:2019tjn,He:2019ify} thought $J^p=\frac{1}{2}^-$ and $J^p=\frac{3}{2}^-$ are compatible when taking both $P_c(4440)^+$ and $P_c(4457)^+$ as the $\Sigma_c\bar{D}^*$ bound state.

In the literature, people attempted to investigate the hidden charm pentaquark system with four quarks and one anti-quark by different techniques: say meson-baryon molecular models where the energy spectrum was calculated by the chiral quark model \cite{Wang:2011rga}, the coupled channel unitary approach \cite{Wu:2010jy,Wu:2012md}, the chiral effective Lagrangian approach \cite{Yang:2011wz,Chen:2015loa}, the QCD sum rules \cite{Chen:2015moa}, the color-screen potential model \cite{Huang:2015uda}, the scattering amplitudes approach \cite{Roca:2015dva};
the diquark-diquark-antiquark model \cite{Maiani:2015vwa,Anisovich:2015cia,Li:2015gta,Ghosh:2015ksa,Wang:2015epa}, and
the compact diquark-triquark model \cite{Lebed:2015tna,Zhu:2015bba}.
In these approaches, for the sakes of simplicity and feasibility the five-body interaction is usually reduced to quasi two-body or three-body cluster interactions. In reality, in fact, there is no priori and sound justification for the establishment of these clusters in multiquark system. In other words, the practical configurations of multiquark systems are still an open question.

In this paper, we  study the spectra of hidden heavy flavor\footnote{Throughout the paper, the heavy flavors mean only the charm and beauty quarks. The top quark mostly decays before forming a bound state, therefore will not be considered. The light flavors here mean up and down quarks, of which the isospin symmetry is satisfied.} pentaquarks, doubly heavy flavor pentaquarks, hidden heavy flavor tetraquarks and doubly heavy flavor tetraquarks in non-relativistic constituent quark model. Since it is hard to find the exact solution for the Schr\"odinger equation of multi-body system and considering the heavy quark masses are much larger than the light ones, in the calculation we assume that within the pentaquark system the light quarks move fast around the two heavy quarks, similar to the situation of a hydrogen molecule.

The paper is organized as  followed. In Sec.~\ref{II}, we give out the Schr\"odinger equations of the multi-body systems.
By virtue of the variational method, with test functions we obtain the optimal values for the ``free" variables. In Sec.~\ref{III}, we calculate the spectra of hidden heavy flavor pentaquarks, doubly heavy flavor pentaquarks, hidden heavy flavor tetraquarks and doubly heavy flavor tetraquarks. Promising decay channels for observation of those multiquark states are analyzed. The last section is left for summary and conclusions.

\section{formulae\label{II}}

\subsection{Five-quark system}
In non-relativistic constituent quark model, the Hamitonian operator for the hidden charm pentaquark ($c\bar{c}qq'q''$) or the doubly charm pentaquark ($ccqq'\bar{q''}$) can be written as (natural units $\hbar=1,~c=1$ implied)

%----------------------
\begin{eqnarray}
%----------------------
\hat{{\mathcal H}}&=&\sum_{i=1}^5 (m_i-\frac{\nabla^2_i}{2m_i})+\sum_{j>i=1}^5 \frac{\lambda_i\cdot\lambda_j }{4} \frac{\alpha_s}{r_{ij}}
\nonumber\\&&
+\sum_{j>i=1}^5 \lambda_i\cdot\lambda_j(b_1 r_{ij}-b_0)+V_S(\mathbf{r_{ij}})+V_L(\mathbf{r_{ij}})
 \, ¡£~~
%----------------------
\label{ham}
%----------------------
\end{eqnarray}
%----------------------
Here, $m_i$ and $r_i$ denote the mass and the position vector of quark $i$, while $r_{ij}=|\mathbf{r_{i}}-\mathbf{r_{j}}|$ is the distance between two quarks; $\lambda_i$ is the Gell-mann matrix of SU(3) color group; $\nabla^2_i$ represents the Laplace operator while $\alpha_s$ is the strong coupling constant. Of this Hamitonian, the first term includes the mass and kinetic energy of individual quarks; the second term shows the color Coulomb interaction; the third one is the color linear confining term with an unknown coefficients $b_i$; the fourth and the fifth ones are spin-dependent and orbital excited terms as in Refs.~\cite{Jaffe:2004ph,Maiani:2004vq,Ali:2011ug,Xing:2018bqt,Zhu:2015bba}. Therein, the spin-dependent term can be expressed as
\begin{eqnarray}
%----------------------
V_S(\mathbf{r_{ij}})&=&
\sum_{j>i=1}^5 (-\frac{3}{8})\frac{C^{ij}}{m_i m_j}\lambda_i\cdot\lambda_j \vec{s}_i\cdot \vec{s}_j v_r\delta(r_{ij})\,,\label{Mh}
%----------------------
\label{ham}
%----------------------
\end{eqnarray}
where $\vec{s}_i=\vec{\sigma}_i/2$ denotes the quark spin operator with Pauli matrix $\vec{\sigma}_i$.

The Schr\"{o}dinger equation for the multi-quark states goes as
\begin{eqnarray}
%----------------------
\hat{{\mathcal H}}\Psi(\mathbf{r_i})&=&E\Psi(\mathbf{r_i})
 \,,
%----------------------
\label{scheq}
%----------------------
\end{eqnarray}
of which the exact solution is not ready and appears to be a tough issue. Considering the spin-dependent term is usually treated as the source of the hyperfine splitting, it plays less influence to calculate the ground state. Meanwhile, the orbital term also relates to the excited states. As for the color linear confinement term, we do know much about it and the coefficients $b_i$ still have large variation degree of freedom. We may get the ground states from the first two terms of Hamitonian (\ref{ham}) through the variational method. Taking account of the spin-dependent and orbit dependent terms, we can then obtain the whole spectra of concerned multiquark states.

Considering the heavy quark masses are much larger than those of the light quarks, we will use Born-Oppenheimer approximation for simplification. The total energy for the five quark system will be split into two heavy quark masses, three light quark masses, kinetic and potential energies of the light degree of freedom, and the spin dependent and orbital excited terms. This separation is valid at the leading order in non-relativistic expansion in velocity of heavy quarks, viz., the two heavy quarks are rest. We assume the distance between two heavy quarks is $R$, a free parameter varying in certain range.

Therefore the Hamitonian operator for the kinetic and potential energies of the light quarks reads

%----------------------
\begin{eqnarray}
%----------------------
\hat{{\mathcal H}}_q&=&-\frac{1}{2m_q}(\nabla^2_1+\nabla^2_2+\nabla^2_3)\nonumber\\&&+ \sum_i^3\frac{ \alpha_s}{4} \lambda_i\cdot\lambda_a(\frac{1}{r_{ia}}+
\frac{4}{\alpha_s}(b_1r_{ia}+b_0))
\nonumber\\&&+ \sum_i^3\frac{ \alpha_s}{4} \lambda_i\cdot\lambda_b(\frac{1}{r_{ib}}+
\frac{4}{\alpha_s}(b_1r_{ib}+b_0))
\nonumber\\&&+ \sum_{j>i=1}^3\frac{ \alpha_s}{4} \lambda_i\cdot\lambda_j(\frac{1}{r_{ij}}+
\frac{4}{\alpha_s}(b_1r_{ij}+b_0))
\nonumber\\&&+ \frac{ \alpha_s}{4} \lambda_a\cdot\lambda_b(\frac{1}{R}+
\frac{4}{\alpha_s}(b_1R+b_0))+V_S(\mathbf{r_{ij}})
 \,.
%----------------------
\label{ham3q}
%----------------------
\end{eqnarray}
Here we especially highlight the heavy quarks by letters ``a" and ``b".

 \begin{figure}[th]
\begin{center}
\includegraphics[width=0.32\textwidth]{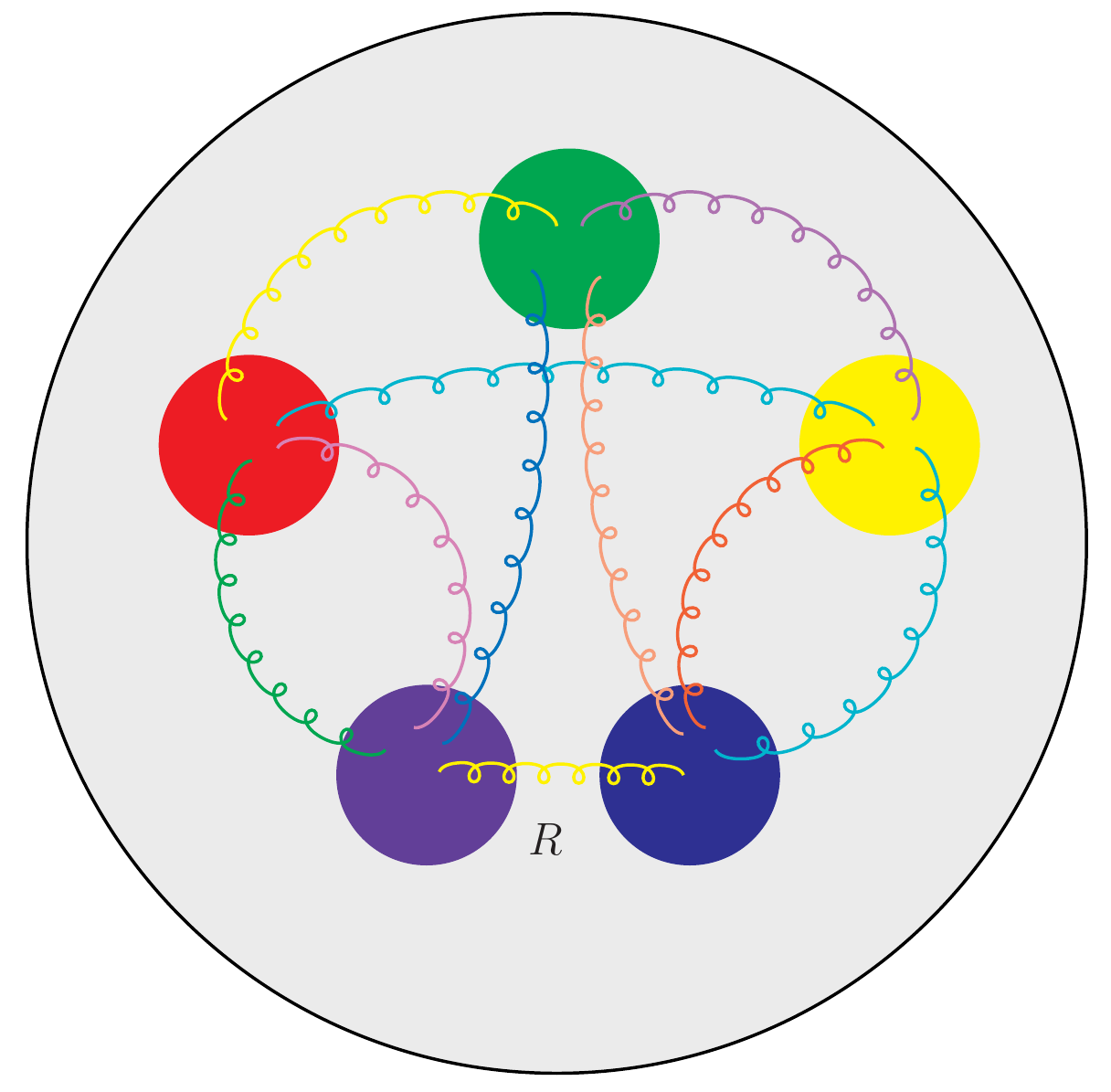}
\caption{The three light quarks orbit around the two heavy quarks, where $R$ is a free parameter representing the distance between two heavy flavors.}\label{tab:pentaquark}
\end{center}
\end{figure}

For pentaquark ($c\bar{c}qq'q''$), where light degrees of freedom orbit around the two rest charm quarks, the wave function can be separated into three sectors, i.e.
 \begin{eqnarray}
\Psi(c\bar{c}qq'q'')&=&\chi_\lambda(\lambda_a,\lambda_b)\otimes \chi_s(s_a,s_b)\otimes \Psi_q(q,q',q'')\,.\nonumber\\
\end{eqnarray}
Here $\Psi_q(q,q',q'')$ is the wave function for the three light quarks, while $\chi_\lambda(\lambda_a,\lambda_b)$ and $\chi_s(s_a,s_b)$ are color and spin wave functions for the heavy quarks, respectively. Note that $\Psi_q(q,q',q'')$ can be decomposed into different sectors, viz. space-coordinate, flavor, color, and spin subspaces,
 \begin{eqnarray}
\Psi_q(q,q',q'')&=&R(\bold{r}_1,\bold{r}_2,\bold{r}_3)\otimes \chi_f(f_1,f_2,f_3)
\nonumber\\&&\otimes \chi_\lambda(\lambda_1,\lambda_2,\lambda_3)\otimes \chi_s(s_1,s_2,s_3)\,,
\end{eqnarray}
where $R(\bold{r}_i)$, $\chi_f(f_i)$,  $\chi_\lambda(\lambda_i)$, and $\chi_s(s_i)$ are the radial, flavor, color, and spin wave functions, respectively. Note, since the isospin symmetry holds in the light degree of freedom, $\chi_f(f_1,f_2,f_3)$ is symmetric under the exchange of two light quarks.

For pentaquark ($c\bar{c}qq'q''$), the color structure of $c\bar{c}$ may be in either $\mathbf{1}_c$ or $\mathbf{8}_c$ representation, so does the color structure of $qq'q''$, since $\mathbf{3}\otimes\mathbf{3}\otimes\mathbf{3}=\mathbf{1}\oplus\mathbf{8}
\oplus\mathbf{8}\oplus\mathbf{10}$. In practice, pentaquark appears in color singlet, and hence $qq'q''$ can not be in the $\mathbf{10}$ configuration.

On the other hand, for pentaquark ($ccqq'\bar{q''}$), the color structure of $cc$ can be in either $\bar{\mathbf{3}}_c$ or $\mathbf{6}_c$ representation, meanwhile the color structure of $qq'\bar{q''}$ would be either in $\mathbf{3}_c$ or $\bar{\mathbf{6}}_c$ representation, because $\mathbf{3}\otimes\mathbf{3}\otimes\bar{\mathbf{3}}=\mathbf{3}\oplus\mathbf{3}
\oplus\bar{\mathbf{6}}\oplus\mathbf{15}$. In this case, the $qq'q''$ can not take the $\mathbf{15}$ configuration.

The color bases for the pentaquark ($c\bar{c}qq'q''$) then appear as $|\mathbf{1}_c^{c\bar{c}}\mathbf{1}_c^{qq'q''}\rangle$ and $|\mathbf{8}_c^{c\bar{c}}\mathbf{8}_c^{qq'q''}\rangle$. It is not unique for the color bases of the pentaquark ($c\bar{c}qq'q''$). However, other structures, like $|\bar{\mathbf{3}}_c^{cq}\mathbf{3}_c^{\bar{c}q'q''}\rangle$ and $|\mathbf{6}_c^{cq}\bar{\mathbf{6}}_c^{\bar{c}q'q''}\rangle$, may also exist.

\begin{table*}[ht]
\renewcommand\arraystretch{1.5}
\caption{\label{color.1}Color matrix elements  for five quark system $[\bar{c}(1)c(2)][q(3) q'(4) q''(5)]$. Therein the superscripts $1$s and $2$s denoting the two-quark color structures are symmetric and antisymmetric, respectively. }
\begin{tabular}{lccccccccccc}
\toprule[1pt] \toprule[1pt]
        &$\hat{O}$~~  &~~$\vec{\lambda_{1}}\cdot\vec{\lambda_{2}}$~~  &~~$\vec{\lambda_{1}}\cdot\vec{\lambda_{3}}$~~ &~~$\vec{\lambda_{1}}\cdot\vec{\lambda_{4}}$~~  &~~$\vec{\lambda_{1}}\cdot\vec{\lambda_{5}}$~~ &~~$\vec{\lambda_{2}}\cdot\vec{\lambda_{3}}$~~ &~~$\vec{\lambda_{2}}\cdot\vec{\lambda_{4}}$~~ &~~$\vec{\lambda_{2}}\cdot\vec{\lambda_{5}}$~~ &~~$\vec{\lambda_{3}}\cdot\vec{\lambda_{4}}$~~ &~~$\vec{\lambda_{3}}\cdot\vec{\lambda_{5}}$~~ &~~$\vec{\lambda_{4}}\cdot\vec{\lambda_{5}}$ \\ \hline

        &~~$<11\mid\hat{O}\mid11>$~~   &~~$-\frac{16}{3}$~~  &~~$0$~~  &~~$0$~~  &~~$0$~~  &~~$0$~~   &~~$0$~~  &~~$0$~~  &~~$-\frac{8}{3}$~~  &~~$-\frac{8}{3}$~~    &~~$-\frac{8}{3}$~~ \\

         &~~$<8^{1}8^{1}\mid\hat{O}\mid8^{1}8^{1}>$~~   &~~$\frac{2}{3}$~~  &~~$-\frac{10}{3}$~~  &~~$-\frac{10}{3}$~~  &~~$\frac{2}{3}$~~  &~~$-\frac{5}{3}$~~   &~~$-\frac{5}{3}$~~  &~~$-\frac{8}{3}$~~   &~~$\frac{4}{3}$~~   &~~$-\frac{5}{3}$~~   &~~$-\frac{5}{3}$~~ \\

          &~~$<8^{2}8^{2}\mid\hat{O}\mid8^{2}8^{2}>$~~   &~~$-1$~~  &~~$-1$~~  &~~$-1$~~  &~~$-1$~~  &~~$-\frac{3}{2}$~~   &~~$-\frac{3}{2}$~~  &~~$-1$~~   &~~$-\frac{5}{3}$~~   &~~$-\frac{1}{6}$~~   &~~$-\frac{3}{2}$~~ \\

        &~~$<11\mid\hat{O}\mid8^{1}8^{1}>$~~   &~~$0$~~  &~~$0$~~  &~~$0$~~  &~~$0$~~  &~~$\sqrt\frac{2}{3}$~~   &~~$-\sqrt\frac{2}{3}$~~  &~~$0$~~   &~~$0$~~   &~~$0$~~   &~~$0$~~ \\

          &~~$<11\mid\hat{O}\mid8^{2}8^{2}>$~~   &~~$0$~~  &~~$0$~~  &~~$0$~~  &~~$0$~~  &~~$-\frac{\sqrt{2}}{3}$~~   &~~$-\frac{\sqrt{2}}{3}$~~  &~~$\frac{2\sqrt{2}}{3}$~~   &~~$0$~~   &~~$0$~~   &~~$0$~~ \\

          &~~$<8^{1}8^{1}\mid\hat{O}\mid8^{2}8^{2}>$~~   &~~$0$~~  &~~$0$~~  &~~$0$~~  &~~$0$~~  &~~$-\frac{1}{2\sqrt{3}}$~~   &~~$\frac{1}{2\sqrt{3}}$~~  &~~$0$~~   &~~$0$~~   &~~$-\frac{\sqrt{3}}{2}$~~   &~~$\frac{\sqrt{3}}{2}$~~ \\

 \\ \bottomrule[1pt]\bottomrule[1pt]

\end{tabular}
\end{table*}

%\begin{sidewaystable}
%\centering
\begin{table*}[ht]
\renewcommand\arraystretch{1.5}
\caption{\label{color.2}Color matrix elements  for five quark system $[c(1)c(2)][q(3)q'(4)]\bar{q}''(5)$.}
\begin{tabular}{lcccccccccccc}
\toprule[1pt] \toprule[1pt]
        &$\hat{O}$~~  &~~$\vec{\lambda_{1}}\cdot\vec{\lambda_{2}}$~~  &~~$\vec{\lambda_{1}}\cdot\vec{\lambda_{3}}$~~ &~~$\vec{\lambda_{1}}\cdot\vec{\lambda_{4}}$~~  &~~$\vec{\lambda_{1}}\cdot\vec{\lambda_{5}}$~~ &~~$\vec{\lambda_{2}}\cdot\vec{\lambda_{3}}$~~ &~~$\vec{\lambda_{2}}\cdot\vec{\lambda_{4}}$~~ &~~$\vec{\lambda_{2}}\cdot\vec{\lambda_{5}}$~~ &~~$\vec{\lambda_{3}}\cdot\vec{\lambda_{4}}$~~ &~~$\vec{\lambda_{3}}\cdot\vec{\lambda_{5}}$~~ &~~$\vec{\lambda_{4}}\cdot\vec{\lambda_{5}}$ \\ \hline

        &~~$<6\bar{3}\bar{3}\mid\hat{O}\mid6\bar{3}\bar{3}>$~~   &~~$\frac{4}{3}$~~  &~~$-\frac{5}{3}$~~  &~~$-\frac{5}{3}$~~  &~~$-\frac{10}{3}$~~  &~~$-\frac{5}{3}$~~   &~~$-\frac{5}{3}$~~  &~~$-\frac{10}{3}$~~   &~~$-\frac{8}{3}$~~   &~~$\frac{2}{3}$~~   &~~$\frac{2}{3}$~~ \\

         &~~$<\bar{3}6\bar{3}\mid\hat{O}\mid\bar{3}6\bar{3}>$~~   &~~$-\frac{8}{3}$~~  &~~$-\frac{5}{3}$~~  &~~$-\frac{5}{3}$~~  &~~$\frac{2}{3}$~~  &~~$-\frac{5}{3}$~~   &~~$-\frac{5}{3}$~~  &~~$\frac{2}{3}$~~   &~~$\frac{4}{3}$~~   &~~$-\frac{10}{3}$~~   &~~$-\frac{10}{3}$~~ \\

          &~~$<\bar{3}\bar{3}\bar{3}\mid\hat{O}\mid\bar{3}\bar{3}\bar{3}>$~~   &~~$-\frac{8}{3}$~~  &~~$-\frac{2}{3}$~~  &~~$-\frac{2}{3}$~~  &~~$-\frac{4}{3}$~~  &~~$-\frac{2}{3}$~~   &~~$-\frac{2}{3}$~~  &~~$-\frac{4}{3}$~~   &~~$-\frac{8}{3}$~~   &~~$-\frac{4}{3}$~~   &~~$-\frac{4}{3}$~~ \\

        &~~$<6\bar{3}\bar{3}\mid\hat{O}\mid\bar{3}6\bar{3}>$~~   &~~$0$~~  &~~$-1$~~ &~~$1$~~ &~~$0$~~    &~~$1$~~   &~~$-1$~~  &~~$0$~~   &~~$0$~~   &~~$0$~~   &~~$0$~~ \\

          &~~$<6\bar{3}\bar{3}\mid\hat{O}\mid\bar{3}\bar{3}\bar{3}>$~~   &~~$0$~~  &~~$-\sqrt{2}$~~  &~~$-\sqrt{2}$~~  &~~$\sqrt{8}$~~  &~~$\sqrt{2}$~~   &~~$\sqrt{2}$~~  &~~$-\sqrt{8}$~~   &~~$0$~~   &~~$0$~~   &~~$0$~~ \\

          &~~$<\bar{3}6\bar{3}\mid\hat{O}\mid\bar{3}\bar{3}\bar{3}>$~~   &~~$0$~~  &~~$\sqrt{2}$~~  &~~$-\sqrt{2}$~~  &~~$0$~~  &~~$\sqrt{2}$~~   &~~$-\sqrt{2}$~~  &~~$0$~~   &~~$0$~~   &~~$-\sqrt{8}$~~   &~~$\sqrt{8}$~~ \\

 \\ \bottomrule[1pt]\bottomrule[1pt]

\end{tabular}
\end{table*}
%\end{sidewaystable}

The color operators can be written as
 \begin{eqnarray}
\lambda_i \cdot \lambda_j&=&
\frac{1}{2}(\lambda^2_{ij}-\lambda^2_{i}-\lambda^2_{j})\,
\end{eqnarray}
with $\lambda^2_{ij}$ being the quadratic Casimir operator and satisfying
 \begin{eqnarray}
\overrightarrow{\lambda}_{i j}^{2} \chi(\lambda \mu)=\frac{4}{3}\left(\lambda^{2}+\mu^{2}+\lambda \mu+3 \lambda+3 \mu\right) \chi(\lambda \mu)\,.
\end{eqnarray}
For the color singlet $\mathbf{1} $,  $\chi(\lambda \mu)=\chi(0 0)$, while for color triplet $\mathbf{3} $,  $\chi(\lambda \mu)=\chi(1 0)$.

The exact solution of the five-body Schr\"{o}dinger equation is not available and hard to get. In the following, we will adopt the variational method with a test  function. The radial wave function of the three light quarks in the ground state of pentaquarks can be assumed as
 \begin{eqnarray}
&&R(\bold{r}_1,\bold{r}_2,\bold{r}_3)=C_s(R_a(\bold{r}_1)R_b(\bold{r}_2)R_a(\bold{r}_3)\nonumber\\
&&~~~~+R_b(\bold{r}_1)R_a(\bold{r}_2)R_a(\bold{r}_3)+R_a(\bold{r}_1)R_a(\bold{r}_2)R_b(\bold{r}_3))\,,\nonumber\\
 \end{eqnarray}
 where the test function is chosen as $R_i(\bold{r}_j)=\frac{\beta^3}{\pi}Exp(-\beta r_{ji})$ and $\beta$ is the free parameter. The normalization constant $C_s$ is expressed as
  \begin{eqnarray}
&&C_s=\sqrt{\frac{1}{3(1+2D^2)}}\,,
 \end{eqnarray}
therein the overlap integral $D$ is defined as
  \begin{eqnarray}
D&=&\int_0^{2\pi}\int_{-1}^{1}\int_0^\infty R_a(\bold{r}_j)R_b(\bold{r}_j) r_j^2 d r_j d\cos\theta d\varphi\nonumber\\&=&
(1+R\beta+\frac{1}{3}R^2\beta^2)Exp(-R\beta)\,.
 \end{eqnarray}

\begin{figure}[th]
\begin{center}
\includegraphics[width=0.32\textwidth]{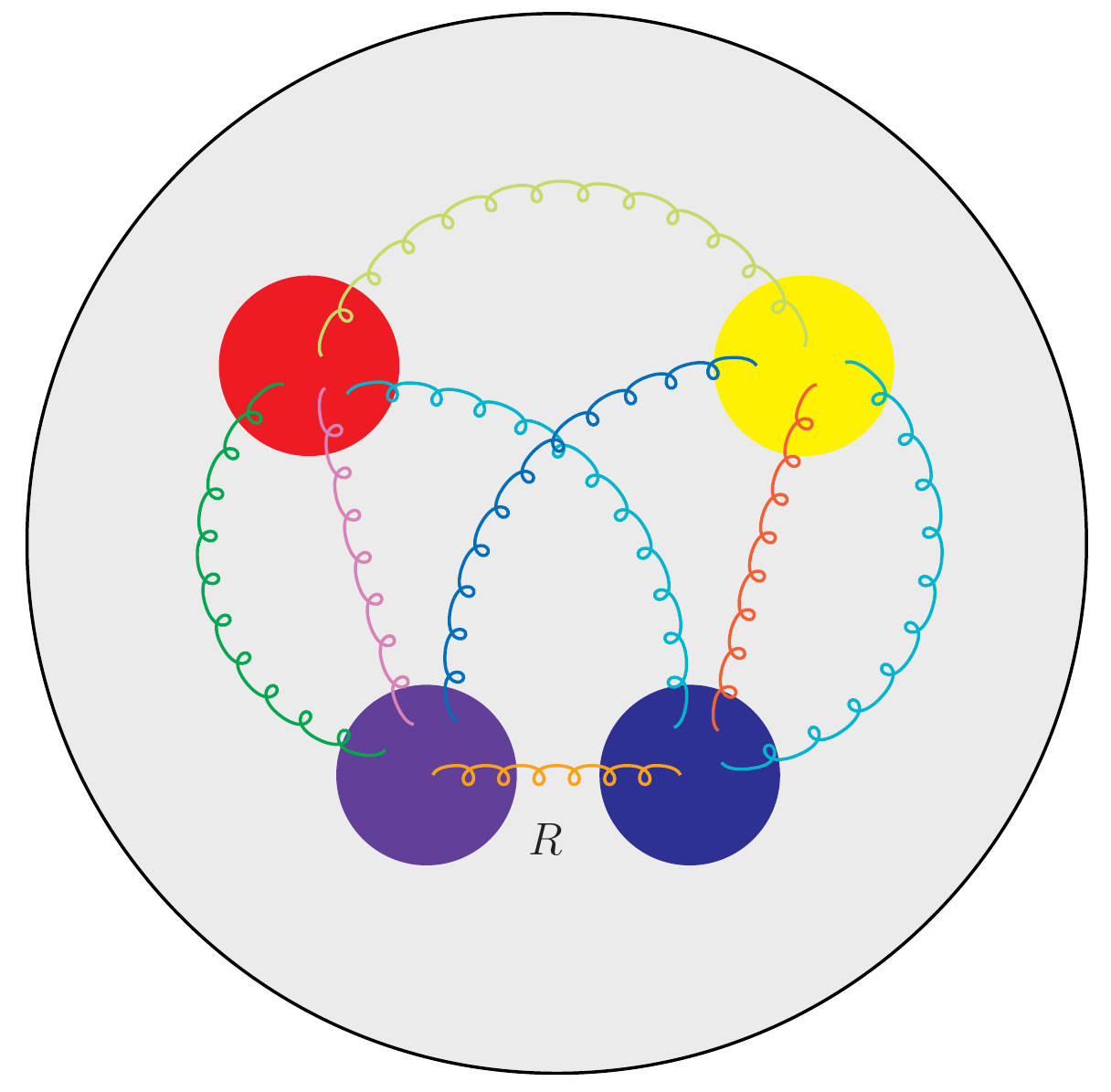}
\caption{Sketchy of the two light quarks orbit around the two heavy flavor quarks. $R$ is the distance between the two heavy quarks. This picture looks like the hydrogen molecule where two electrons move around the two protons. }\label{tab:tetraquark}
\end{center}
\end{figure}

\small
\begin{table*}[ht]
\renewcommand\arraystretch{1.5}
\caption{\label{spin10.5.1}Spin matrix elements  for five quark system $[\bar{c}(1)c(2)][q(3) q'(4) q''(5)]$. Therein the superscripts $1$s and $2$s denoting the two-quark spin structures are symmetric and antisymmetric, respectively.}
\begin{tabular}{lccccccccccccc}
\toprule[1pt] \toprule[1pt]
       &$J^{P}$~~ &$\hat{B}$~~  &$\vec{\sigma_{1}}\cdot\vec{\sigma_{2}}$~~ &$\vec{\sigma_{1}}\cdot\vec{\sigma_{3}}$~~  &$\vec{\sigma_{1}}\cdot\vec{\sigma_{4}}$~~ &$\vec{\sigma_{1}}\cdot\vec{\sigma_{5}}$~~ &$\vec{\sigma_{2}}\cdot\vec{\sigma_{3}}$~~ &$\vec{\sigma_{2}}\cdot\vec{\sigma_{4}}$~~ &$\vec{\sigma_{2}}\cdot\vec{\sigma_{5}}$~~ &$\vec{\sigma_{3}}\cdot\vec{\sigma_{4}}$~~ &$\vec{\sigma_{3}}\cdot\vec{\sigma_{5}}$~~ &$\vec{\sigma_{4}}\cdot\vec{\sigma_{5}}$~~ \\ \hline

       &~~$\frac{1}{2}^{-}$~~ &~~$<1\frac{3}{2}\mid\hat{B}\mid1\frac{3}{2}>$~~   &~~$1$~~  &~~$-\frac{5}{3}$~~  &~~$-\frac{5}{3}$~~  &~~$-\frac{5}{3}$~~  &~~$-\frac{5}{3}$~~   &~~$-\frac{5}{3}$~~  &~~$-\frac{5}{3}$~~   &~~$1$~~   &~~$1$~~   &~~$1$~~ \\
     &~~$$~~ &~~$<0\frac{1}{2}^{1}\mid\hat{B}\mid0\frac{1}{2}^{1}>$~~   &~~$-3$~~  &~~$0$~~  &~~$0$~~  &~~$0$~~  &~~$0$~~   &~~$0$~~  &~~$0$~~   &~~$1$~~   &~~$-2$~~   &~~$-2$~~ \\

  &~~$$~~ &~~$<0\frac{1}{2}^{2}\mid\hat{B}\mid0\frac{1}{2}^{2}>$~~   &~~$-3$~~  &~~$0$~~  &~~$0$~~  &~~$0$~~  &~~$0$~~   &~~$0$~~  &~~$0$~~   &~~$-3$~~   &~~$0$~~   &~~$0$~~ \\

   &~~$$~~ &~~$<1\frac{1}{2}^{1}\mid\hat{B}\mid1\frac{1}{2}^{1}>$~~   &~~$1$~~  &~~$-\frac{4}{3}$~~  &~~$-\frac{4}{3}$~~  &~~$\frac{2}{3}$~~  &~~$-\frac{4}{3}$~~   &~~$-\frac{4}{3}$~~  &~~$\frac{2}{3}$~~   &~~$1$~~   &~~$-2$~~   &~~$-2$~~ \\

    &~~$$~~ &~~$<1\frac{1}{2}^{2}\mid\hat{B}\mid1\frac{1}{2}^{2}>$~~   &~~$1$~~  &~~$0$~~  &~~$0$~~  &~~$-2$~~  &~~$0$~~   &~~$0$~~  &~~$-2$~~   &~~$-3$~~   &~~$0$~~   &~~$0$~~ \\

    &~~$$~~ &~~$<1\frac{3}{2}\mid\hat{B}\mid0\frac{1}{2}^{1}>$~~   &~~$0$~~  &~~$-\sqrt{\frac{2}{3}}$~~  &~~$-\sqrt{\frac{2}{3}}$~~  &~~$\sqrt{\frac{8}{3}}$~~  &~~$\sqrt{\frac{2}{3}}$~~   &~~$\sqrt{\frac{2}{3}}$~~  &~~$-\sqrt{\frac{8}{3}}$~~   &~~$0$~~   &~~$0$~~   &~~$0$~~ \\

    &~~$$~~ &~~$<1\frac{3}{2}\mid\hat{B}\mid0\frac{1}{2}^{2}>$~~   &~~$0$~~  &~~$-\sqrt{2}$~~  &~~$\sqrt{2}$~~  &~~$0$~~  &~~$\sqrt{2}$~~   &~~$-\sqrt{2}$~~  &~~$0$~~   &~~$0$~~   &~~$0$~~   &~~$0$~~ \\

     &~~$$~~ &~~$<1\frac{3}{2}\mid\hat{B}\mid1\frac{1}{2}^{1}>$~~   &~~$0$~~  &~~$-\sqrt{\frac{2}{9}}$~~  &~~$-\sqrt{\frac{2}{9}}$~~  &~~$\sqrt{\frac{8}{9}}$~~  &~~$-\sqrt{\frac{2}{9}}$~~   &~~$-\sqrt{\frac{2}{9}}$~~  &~~$\sqrt{\frac{8}{9}}$~~   &~~$0$~~   &~~$0$~~   &~~$0$~~ \\

     &~~$$~~ &~~$<1\frac{3}{2}\mid\hat{B}\mid1\frac{1}{2}^{2}>$~~   &~~$0$~~  &~~$-\sqrt{\frac{2}{3}}$~~  &~~$\sqrt{\frac{2}{3}}$~~  &~~$0$~~     &~~$-\sqrt{\frac{2}{3}}$~~  &~~$\sqrt{\frac{2}{3}}$~~  &~~$0$~~ &~~$0$~~   &~~$0$~~   &~~$0$~~ \\

     &~~$$~~ &~~$<0\frac{1}{2}^{1}\mid\hat{B}\mid0\frac{1}{2}^{2}>$~~   &~~$0$~~  &~~$0$~~  &~~$0$~~  &~~$0$~~  &~~$0$~~   &~~$0$~~  &~~$0$~~   &~~$0$~~   &~~$-\sqrt{3}$~~   &~~$\sqrt{3}$~~ \\

     &~~$$~~ &~~$<0\frac{1}{2}^{1}\mid\hat{B}\mid1\frac{1}{2}^{1}>$~~   &~~$0$~~  &~~$\sqrt{\frac{4}{3}}$~~  &~~$\sqrt{\frac{4}{3}}$~~  &~~$-\sqrt{\frac{1}{3}}$~~  &~~$-\sqrt{\frac{4}{3}}$~~   &~~$-\sqrt{\frac{4}{3}}$~~  &~~$\sqrt{\frac{1}{3}}$~~   &~~$0$~~   &~~$0$~~   &~~$0$~~ \\

     &~~$$~~ &~~$<0\frac{1}{2}^{1}\mid\hat{B}\mid1\frac{1}{2}^{2}>$~~   &~~$0$~~  &~~$-1$~~  &~~$1$~~  &~~$0$~~  &~~$1$~~   &~~$-1$~~  &~~$0$~~   &~~$0$~~   &~~$0$~~   &~~$0$~~ \\

      &~~$$~~ &~~$<0\frac{1}{2}^{2}\mid\hat{B}\mid1\frac{1}{2}^{1}>$~~   &~~$0$~~  &~~$-1$~~  &~~$1$~~  &~~$0$~~  &~~$1$~~   &~~$-1$~~  &~~$0$~~   &~~$0$~~   &~~$0$~~   &~~$0$~~ \\

      &~~$$~~ &~~$<0\frac{1}{2}^{2}\mid\hat{B}\mid1\frac{1}{2}^{2}>$~~   &~~$0$~~  &~~$0$~~  &~~$0$~~  &~~$\sqrt{3}$~~  &~~$0$~~   &~~$0$~~  &~~$-\sqrt{3}$~~   &~~$0$~~   &~~$0$~~   &~~$0$~~ \\

       &~~$$~~ &~~$<1\frac{1}{2}^{1}\mid\hat{B}\mid1\frac{1}{2}^{2}>$~~   &~~$0$~~  &~~$\sqrt{\frac{4}{3}}$~~  &~~$-\sqrt{\frac{4}{3}}$~~  &~~$0$~~  &~~$\sqrt{\frac{4}{3}}$~~   &~~$-\sqrt{\frac{4}{3}}$~~  &~~$0$~~   &~~$0$~~   &~~$-\sqrt{3}$~~   &~~$\sqrt{3}$~~ \\

       &~~$\frac{3}{2}^{-}$~~ &~~$<0\frac{3}{2}\mid\hat{B}\mid0\frac{3}{2}>$~~   &~~$-3$~~  &~~$0$~~  &~~$0$~~  &~~$0$~~  &~~$0$~~   &~~$0$~~  &~~$0$~~   &~~$1$~~   &~~$1$~~   &~~$1$~~ \\

       &~~$$~~ &~~$<1\frac{3}{2}\mid\hat{B}\mid1\frac{3}{2}>$~~   &~~$1$~~  &~~$-\frac{2}{3}$~~  &~~$-\frac{2}{3}$~~  &~~$-\frac{1}{3}$~~  &~~$-\frac{2}{3}$~~   &~~$-\frac{2}{3}$~~  &~~$-\frac{1}{3}$~~   &~~$1$~~   &~~$1$~~   &~~$1$~~ \\

       &~~$$~~ &~~$<1\frac{1}{2}^{1}\mid\hat{B}\mid1\frac{1}{2}^{1}>$~~   &~~$1$~~  &~~$\frac{2}{3}$~~  &~~$\frac{2}{3}$~~  &~~$-\frac{1}{3}$~~  &~~$\frac{2}{3}$~~   &~~$\frac{2}{3}$~~  &~~$-\frac{1}{3}$~~   &~~$1$~~   &~~$1$~~   &~~$1$~~ \\

       &~~$$~~ &~~$<1\frac{1}{2}^{2}\mid\hat{B}\mid1\frac{1}{2}^{2}>$~~   &~~$1$~~  &~~$0$~~  &~~$0$~~  &~~$1$~~  &~~$0$~~   &~~$0$~~  &~~$1$~~   &~~$1$~~   &~~$-2$~~   &~~$-2$~~ \\

       &~~$$~~ &~~$<0\frac{3}{2}\mid\hat{B}\mid1\frac{3}{2}>$~~   &~~$0$~~  &~~$\sqrt{\frac{5}{3}}$~~  &~~$\sqrt{\frac{5}{3}}$~~  &~~$\sqrt{\frac{5}{3}}$~~  &~~$-\sqrt{\frac{5}{3}}$~~   &~~$-\sqrt{\frac{5}{3}}$~~  &~~$-\sqrt{\frac{5}{3}}$~~   &~~$-3$~~   &~~$0$~~   &~~$0$~~ \\

       &~~$$~~ &~~$<0\frac{3}{2}\mid\hat{B}\mid1\frac{1}{2}^{1}>$~~   &~~$0$~~  &~~$\sqrt{\frac{1}{3}}$~~  &~~$\sqrt{\frac{1}{3}}$~~ &~~$-\sqrt{\frac{4}{3}}$~~  &~~$-\sqrt{\frac{1}{3}}$~~   &~~$-\sqrt{\frac{1}{3}}$~~  &~~$\sqrt{\frac{4}{3}}$~~   &~~$0$~~   &~~$0$~~   &~~$0$~~ \\

       &~~$$~~ &~~$<0\frac{3}{2}\mid\hat{B}\mid1\frac{1}{2}^{2}>$~~   &~~$0$~~  &~~$1$~~  &~~$-1$~~  &~~$0$~~  &~~$-1$~~   &~~$1$~~  &~~$0$~~   &~~$0$~~   &~~$0$~~   &~~$0$~~ \\

       &~~$$~~ &~~$<1\frac{3}{2}\mid\hat{B}\mid1\frac{1}{2}^{1}>$~~   &~~$0$~~  &~~$-\sqrt{\frac{5}{9}}$~~  &~~$-\sqrt{\frac{5}{9}}$~~  &~~$\sqrt{\frac{20}{9}}$~~  &~~$-\sqrt{\frac{5}{9}}$~~   &~~$-\sqrt{\frac{5}{9}}$~~  &~~$\sqrt{\frac{20}{9}}$~~   &~~$0$~~   &~~$0$~~   &~~$0$~~ \\

       &~~$$~~ &~~$<1\frac{3}{2}\mid\hat{B}\mid1\frac{1}{2}^{2}>$~~   &~~$0$~~  &~~$-\sqrt{\frac{5}{3}}$~~  &~~$\sqrt{\frac{5}{3}}$~~  &~~$0$~~  &~~$-\sqrt{\frac{5}{3}}$~~   &~~$\sqrt{\frac{5}{3}}$~~  &~~$0$~~   &~~$0$~~   &~~$0$~~   &~~$0$~~ \\

      &~~$$~~ &~~$<1\frac{1}{2}^{1}\mid\hat{B}\mid1\frac{1}{2}^{2}>$~~   &~~$0$~~  &~~$-\sqrt{\frac{1}{3}}$~~  &~~$\sqrt{\frac{1}{3}}$~~  &~~$0$~~ &~~$-\sqrt{\frac{1}{3}}$~~   &~~$\sqrt{\frac{1}{3}}$~~  &~~$0$~~   &~~$0$~~   &~~$-\sqrt{3}$~~   &~~$\sqrt{3}$~~ \\

       &~~$\frac{5}{2}^{-}$~~ &~~$<1\frac{3}{2}\mid\hat{B}\mid1\frac{3}{2}>$~~   &~~$1$~~  &~~$1$~~  &~~$1$~~  &~~$1$~~ &~~$1$~~   &~~$1$~~  &~~$1$~~   &~~$1$~~   &~~$1$~~   &~~$1$~~ \\

 \\ \bottomrule[1pt]\bottomrule[1pt]

\end{tabular}
\end{table*}

\subsection{Four quarks system}

Similarly, the total energy of the four quark system may be decomposed as two heavy quark masses, two light quark masses, the kinetic and potential energies of the two light quarks, and the spin dependent and orbital excited terms. The Hamitonian for the kinetic and potential energies of the two light quarks with trivial orbital angular momentum reads
%----------------------
\begin{eqnarray}
%----------------------
\hat{{\mathcal H}}'_q&=&-\frac{1}{2m_q}(\nabla^2_1+\nabla^2_2)\nonumber\\&&+ \sum_i^2\frac{ \alpha_s}{4} \lambda_i\cdot\lambda_a(\frac{1}{r_{ia}}+
\frac{4}{\alpha_s}(b_1r_{ia}+b_0))
\nonumber\\&&+ \sum_i^2\frac{ \alpha_s}{4} \lambda_i\cdot\lambda_b(\frac{1}{r_{ib}}+
\frac{4}{\alpha_s}(b_1r_{ib}+b_0))
\nonumber\\&&+ \frac{ \alpha_s}{4} \lambda_1\cdot\lambda_2(\frac{1}{r_{12}}+
\frac{4}{\alpha_s}(b_1r_{12}+b_0))
\nonumber\\&&+ \frac{ \alpha_s}{4} \lambda_a\cdot\lambda_b(\frac{1}{R}+
\frac{4}{\alpha_s}(b_1R+b_0))+V_S(\mathbf{r_{ij}})
 \,.~~~
%----------------------
\label{ham3q}
%----------------------
\end{eqnarray}

For the tetraquark ($\bar{c}\bar{c} qq'$) where  two light quarks orbit around the two  rest charm quarks, the wave function can be separated into two part
 \begin{eqnarray}
\Psi(\bar{c}\bar{c}qq')&=&\chi_\lambda(\lambda_a,\lambda_b)\otimes \chi_s(s_a,s_b)\otimes \Psi_q(q,q')\,,\nonumber\\
\end{eqnarray}
where
 \begin{eqnarray}
\Psi_q(q,q')&=&R(\bold{r}_1,\bold{r}_2)\otimes \chi_f(f_1,f_2)
\nonumber\\&&\otimes \chi_\lambda(\lambda_1,\lambda_2)\otimes \chi_s(s_1,s_2)\,,
\end{eqnarray}
Due to the Isospin symmetry for the light quarks, $\chi_f(f_1,f_2)$ is also symmetrical.

For the Tetraquark ($\bar{c}\bar{c} qq'$), the color structure of $\bar{c}\bar{c}$ is $\mathbf{3}_c$ or $\bar{\mathbf{6}}_c$ representation, while the color structure of $qq'$ is  $\bar{\mathbf{3}}_c$ or $\mathbf{6}_c$ representation. Considering the anti-symmetrical properties for identical fermions, the
spin quantum number of $\bar{c}\bar{c}$ is 0 for color triplet and 1 for color anti-sextet, while the
spin quantum number of $qq'$ is 0 for color anti-triplet and 1 for color sextet.
For the Tetraquark ($c \bar{c} q\bar{q'}$), the color structure  is $\mathbf{1}_c$ or $\mathbf{8}_c$ representation for both $c\bar{c}$ and $q\bar{q'}$.

The  strict solution for the four-body Schr\"{o}dinger equation is also not clear.  The radial wave function of the two light quarks in the ground state of pentaquarks can be assumed as
 \begin{eqnarray}
&&R(\bold{r}_1,\bold{r}_2)=C'_s(R_a(\bold{r}_1)R_b(\bold{r}_2))+R_b(\bold{r}_1)R_a(\bold{r}_2))\,,~~
 \end{eqnarray}
 where the normalization constant $C'_s$ is expressed as
  \begin{eqnarray}
&&C'_s=\sqrt{\frac{1}{2(1+D^2)}}\,,
 \end{eqnarray}

\small
\begin{table*}[ht]
\renewcommand\arraystretch{1.5}
\caption{\label{spin9.5.1}Spin matrix elements for five quark system $[c(1)c(2)][q(3)q'(4)]\bar{q}''(5)$.}
\begin{tabular}{lcccccccccccccc}
\toprule[1pt] \toprule[1pt]
       &$J^{P}$~~  &$S_{ccqq}$~~  &$\hat{B}$~~  &$\vec{\sigma_{1}}\cdot\vec{\sigma_{2}}$~~ &$\vec{\sigma_{1}}\cdot\vec{\sigma_{3}}$~~  &$\vec{\sigma_{1}}\cdot\vec{\sigma_{4}}$~~ &$\vec{\sigma_{1}}\cdot\vec{\sigma_{5}}$~~ &$\vec{\sigma_{2}}\cdot\vec{\sigma_{3}}$~~ &$\vec{\sigma_{2}}\cdot\vec{\sigma_{4}}$~~ &$\vec{\sigma_{2}}\cdot\vec{\sigma_{5}}$~~ &$\vec{\sigma_{3}}\cdot\vec{\sigma_{4}}$~~ &$\vec{\sigma_{3}}\cdot\vec{\sigma_{5}}$~~ &$\vec{\sigma_{4}}\cdot\vec{\sigma_{5}}$~~\\ \hline

       &~~$\frac{1}{2}^{-}$~~ &~~$0$~~  &~~$<11\frac{1}{2}\mid\hat{B}\mid11\frac{1}{2}>$~~   &~~$1$~~  &~~$-\frac{2}{3}$~~  &~~$-\frac{2}{3}$~~  &~~$0$~~  &~~$-\frac{2}{3}$~~   &~~$-\frac{2}{3}$~~  &~~$0$~~   &~~$1$~~   &~~$0$~~   &~~$0$~~ \\

       &~~$$~~ &~~$1$~~  &~~$<01\frac{1}{2}\mid\hat{B}\mid01\frac{1}{2}>$~~   &~~$-3$~~  &~~$0$~~  &~~$0$~~  &~~$0$~~  &~~$0$~~   &~~$0$~~  &~~$0$~~   &~~$1$~~   &~~$-2$~~   &~~$-2$~~ \\

       &~~$$~~ &~~$$~~  &~~$<10\frac{1}{2}\mid\hat{B}\mid10\frac{1}{2}>$~~   &~~$1$~~  &~~$0$~~  &~~$0$~~  &~~$-2$~~  &~~$0$~~   &~~$0$~~  &~~$-2$~~   &~~$-3$~~   &~~$0$~~   &~~$0$~~ \\

       &~~$$~~ &~~$$~~  &~~$<11\frac{1}{2}\mid\hat{B}\mid11\frac{1}{2}>$~~   &~~$1$~~  &~~$-1$~~  &~~$-1$~~  &~~$-1$~~  &~~$-1$~~   &~~$-1$~~  &~~$-1$~~   &~~$1$~~   &~~$-1$~~   &~~$-1$~~ \\

       &~~$$~~ &~~$$~~  &~~$<01\frac{1}{2}\mid\hat{B}\mid10\frac{1}{2}>$~~   &~~$0$~~  &~~$1$~~  &~~$-1$~~  &~~$0$~~  &~~$-1$~~   &~~$1$~~  &~~$0$~~   &~~$0$~~   &~~$0$~~   &~~$0$~~ \\

       &~~$$~~ &~~$$~~  &~~$<01\frac{1}{2}\mid\hat{B}\mid11\frac{1}{2}>$~~ &~~$0$~~   &~~$-\sqrt{2}$~~  &~~$-\sqrt{2}$~~  &~~$\sqrt{2}$~~  &~~$\sqrt{2}$~~  &~~$\sqrt{2}$~~   &~~$-\sqrt{2}$~~    &~~$0$~~   &~~$0$~~   &~~$0$~~ \\

       &~~$$~~ &~~$$~~  &~~$<10\frac{1}{2}\mid\hat{B}\mid11\frac{1}{2}>$~~ &~~$0$~~   &~~$\sqrt{2}$~~  &~~$-\sqrt{2}$~~  &~~$0$~~  &~~$\sqrt{2}$~~  &~~$-\sqrt{2}$~~   &~~$0$~~    &~~$0$~~   &~~$-\sqrt{2}$~~   &~~$\sqrt{2}$~~ \\

       &~~$\frac{3}{2}^{-}$~~ &~~$1$~~  &~~$<01\frac{1}{2}\mid\hat{B}\mid01\frac{1}{2}>$~~ &~~$-3$~~   &~~$0$~~  &~~$0$~~  &~~$0$~~  &~~$0$~~  &~~$0$~~   &~~$0$~~    &~~$1$~~   &~~$1$~~   &~~$1$~~ \\

       &~~$$~~ &~~$$~~  &~~$<10\frac{1}{2}\mid\hat{B}\mid10\frac{1}{2}>$~~ &~~$1$~~   &~~$0$~~  &~~$0$~~  &~~$1$~~  &~~$0$~~  &~~$0$~~   &~~$1$~~   &~~$-3$~~   &~~$0$~~ &~~$0$~~\\

       &~~$$~~ &~~$$~~  &~~$<11\frac{1}{2}\mid\hat{B}\mid11\frac{1}{2}>$~~ &~~$1$~~   &~~$-1$~~  &~~$-1$~~  &~~$\frac{1}{2}$~~  &~~$-1$~~  &~~$-1$~~   &~~$\frac{1}{2}$~~    &~~$1$~~   &~~$\frac{1}{2}$~~   &~~$\frac{1}{2}$~~ \\

       &~~$$~~ &~~$$~~  &~~$<01\frac{1}{2}\mid\hat{B}\mid10\frac{1}{2}>$~~ &~~$0$~~   &~~$1$~~  &~~$-1$~~  &~~$0$~~  &~~$-1$~~  &~~$1$~~   &~~$0$~~    &~~$0$~~   &~~$0$~~   &~~$0$~~ \\

       &~~$$~~ &~~$$~~  &~~$<01\frac{1}{2}\mid\hat{B}\mid11\frac{1}{2}>$~~ &~~$0$~~   &~~$-\sqrt{2}$~~  &~~$-\sqrt{2}$~~  &~~$-\sqrt{\frac{1}{2}}$~~   &~~$\sqrt{2}$~~   &~~$\sqrt{2}$~~    &~~$\sqrt{\frac{1}{2}}$~~   &~~$0$~~   &~~$0$~~  &~~$0$~~\\

       &~~$$~~ &~~$$~~  &~~$<10\frac{1}{2}\mid\hat{B}\mid11\frac{1}{2}>$~~ &~~$0$~~   &~~$\sqrt{2}$~~  &~~$-\sqrt{2}$~~  &~~$0$~~  &~~$\sqrt{2}$~~  &~~$-\sqrt{2}$~~ &~~$0$~~ &~~$0$~~ &~~$\sqrt{\frac{1}{2}}$~~ &~~$-\sqrt{\frac{1}{2}}$~~ \\

       &~~$$~~ &~~$2$~~  &~~$<11\frac{1}{2}\mid\hat{B}\mid11\frac{1}{2}>$~~ &~~$1$~~   &~~$1$~~  &~~$1$~~  &~~$-\frac{3}{2}$~~  &~~$1$~~  &~~$1$~~   &~~$-\frac{3}{2}$~~    &~~$1$~~   &~~$-\frac{3}{2}$~~   &~~$-\frac{3}{2}$~~ \\

       &~~$\frac{5}{2}^{-}$~~ &~~$2$~~  &~~$<11\frac{1}{2}\mid\hat{B}\mid11\frac{1}{2}>$~~ &~~$1$~~   &~~$1$~~  &~~$1$~~  &~~$1$~~  &~~$1$~~  &~~$1$~~   &~~$1$~~    &~~$1$~~   &~~$1$~~   &~~$1$~~ \\

 \\ \bottomrule[1pt]\bottomrule[1pt]

\end{tabular}
\end{table*}

\section{Results and discusstion\label{III}}

Combining the spin parts, the spin-color bases for the pentaquark ($[c\bar{c}][qq']q''$) can be written as

  \begin{eqnarray}
&&|0_{cc},1_{qq'},{\frac{1}{2}}_{q''}\rangle_s \otimes|\mathbf{1}_{c\bar{c}},\bar{\mathbf{3}}_{qq'},\mathbf{3}_{q''}\rangle_c\,,\nonumber\\
&&|1_{cc},1_{qq'},{\frac{1}{2}}_{q''}\rangle_s \otimes|\mathbf{1}_{c\bar{c}},\bar{\mathbf{3}}_{qq'},\mathbf{3}_{q''}\rangle_c\,,\nonumber\\
&&|0_{cc},0_{qq'},{\frac{1}{2}}_{q''}\rangle_s \otimes|\mathbf{8}_{c\bar{c}},\mathbf{6}_{qq'},\mathbf{3}_{q''}\rangle_c\,,\nonumber\\
&&|1_{cc},0_{qq'},{\frac{1}{2}}_{q''}\rangle_s \otimes|\mathbf{8}_{c\bar{c}},\mathbf{6}_{qq'},\mathbf{3}_{q''}\rangle_c\,,\nonumber\\
&&|0_{cc},1_{qq'},{\frac{1}{2}}_{q''}\rangle_s \otimes|\mathbf{8}_{c\bar{c}},\bar{\mathbf{3}}_{qq'},\mathbf{3}_{q''}\rangle_c\,,\nonumber\\
&&|1_{cc},1_{qq'},{\frac{1}{2}}_{q''}\rangle_s \otimes|\mathbf{8}_{c\bar{c}},\bar{\mathbf{3}}_{qq'},\mathbf{3}_{q''}\rangle_c\,.
 \end{eqnarray}

Similarly, the spin-color bases for the pentaquark ($[cc][qq']\bar{q''}$), tetraquark ($[c\bar{c}][q\bar{q'}]$, and tetraquark ($[\bar{c}\bar{c}][qq']$) can be written as
  \begin{eqnarray}
&&|0_{cc},1_{qq'},{\frac{1}{2}}_{\bar{q''}}\rangle_s \otimes|\mathbf{6}_{cc},\bar{\mathbf{3}}_{qq'},\bar{\mathbf{3}}_{\bar{q}''}\rangle_c\,,\nonumber\\
&&|1_{cc},0_{qq'},{\frac{1}{2}}_{\bar{q''}}\rangle_s \otimes|\bar{\mathbf{3}}_{cc},\mathbf{6}_{qq'},\bar{\mathbf{3}}_{\bar{q}''}\rangle_c\,,\nonumber\\
&&|1_{cc},1_{qq'},{\frac{1}{2}}_{\bar{q''}}\rangle_s \otimes|\bar{\mathbf{3}}_{cc},\bar{\mathbf{3}}_{qq'},\bar{\mathbf{3}}_{\bar{q}''}\rangle_c\, ,
 \end{eqnarray}
  \begin{eqnarray}
&&|0_{c\bar{c}},0_{q\bar{q'}}\rangle_s \otimes|\mathbf{1}_{c\bar{c}},\bar{\mathbf{1}}_{q\bar{q'}}\rangle_c (\mathrm{or}~ |\mathbf{8}_{c\bar{c}},\bar{\mathbf{8}}_{q\bar{q'}}\rangle_c)\,,\nonumber\\
&&|0_{c\bar{c}},1_{q\bar{q'}}\rangle_s \otimes|\mathbf{1}_{c\bar{c}},\bar{\mathbf{1}}_{q\bar{q'}}\rangle_c (\mathrm{or}~ |\mathbf{8}_{c\bar{c}},\bar{\mathbf{8}}_{q\bar{q'}}\rangle_c)\,,\nonumber\\
&&|1_{c\bar{c}},0_{q\bar{q'}}\rangle_s \otimes|\mathbf{1}_{c\bar{c}},\bar{\mathbf{1}}_{q\bar{q'}}\rangle_c (\mathrm{or}~ |\mathbf{8}_{c\bar{c}},\bar{\mathbf{8}}_{q\bar{q'}}\rangle_c)\,,\nonumber\\
&&|1_{c\bar{c}},1_{q\bar{q'}}\rangle_s \otimes|\mathbf{1}_{c\bar{c}},\bar{\mathbf{1}}_{q\bar{q'}}\rangle_c (\mathrm{or}~ |\mathbf{8}_{c\bar{c}},\bar{\mathbf{8}}_{q\bar{q'}}\rangle_c)\, ,
 \end{eqnarray}
and
\begin{eqnarray}
&&|0_{\bar{c}\bar{c}},0_{qq'}\rangle_s \otimes|\bar{\mathbf{6}}_{\bar{c}\bar{c}},\mathbf{6}_{qq'}\rangle_c\,,\nonumber\\
&&|1_{\bar{c}\bar{c}},1_{qq'}\rangle_s \otimes|\mathbf{3}_{\bar{c}\bar{c}},\bar{\mathbf{3}}_{qq'}\rangle_c\, ,
 \end{eqnarray}
respectively.

%\begin{sidewaystable}
%\centering
\begin{table*}[ht]
\renewcommand\arraystretch{1.5}
\caption{\label{spectra.p} Hidden- and double-charm pentaquark spectra (in GeV) under the classification of the heavy quark spin symmetry. The $s_\ell$ in the table represents the spin of the light degrees of freedom with $\hat{s_\ell}=\hat{J}-\hat{s}_{HQP}$ where the heavy quark pair spin $s_{HQP}$ may be either $s_{c\bar{c}}$ or $s_{cc}$. The parameters given in Set II are widely used in the literature and are also employed in our analysis for pentaquarks. The uncertainty in the table stems from the variation of parameters. It should be mentioned that another source of uncertainty is not included here, i.e. the standard deviation $\sigma=\sqrt{\langle \hat{{\mathcal H}}^2\rangle-\langle \hat{{\mathcal H}}\rangle^2}$ from the test function, which may induce about $\pm100$ MeV uncertainties to the final results. Note, when the pentaquark ground state with certain $J^P$ is above the threshold of constituent quark masses, it will be highlighted with an asterisk.}
\begin{tabular}{lc|c|c|c|c|c|c|c}
\toprule[1pt] \toprule[1pt]
      &Constituents~~   & Color structure &Mass~~    &~~$J^P$~~   &~~$s_{c\bar{c}}$ or $s_{cc}$~~ &~~$s_\ell$~~& Multiplet or singlet &~~Label~~ \\ \hline
& $c\bar{c}qq'q''$ &$|\mathbf{1}_{c\bar{c}},\bar{\mathbf{3}}_{qq'},\mathbf{3}_{q''}\rangle$& $4.269\pm0.059$~~  &~~$\frac{1}{2}^-$~~   &~~0~~ &~~$\frac{1}{2}$~~ & Triplet & 1\\
& $c\bar{c}qq'q''$ &$|\mathbf{8}_{c\bar{c}},\mathbf{6}_{qq'},\mathbf{3}_{q''}\rangle$& ${4.593^*}\pm0.088$~~  &~~$\frac{1}{2}^-$~~   &~~0~~ &~~$\frac{1}{2}$~~ & Triplet & 2\\
& $c\bar{c}qq'q''$ &$|\mathbf{8}_{c\bar{c}},\bar{\mathbf{3}}_{qq'},\mathbf{3}_{q''}\rangle$& $4.468^{+0.076}_{-0.075}$~~  &~~$\frac{1}{2}^-$~~   &~~0~~ &~~$\frac{1}{2}$~~ & Triplet & 3\\\hline
& $c\bar{c}qq'q''$ &$|\mathbf{1}_{c\bar{c}},\bar{\mathbf{3}}_{qq'},\mathbf{3}_{q''}\rangle$& ${4.559^*} ^{+0.085}_{-0.086}$~~  &~~$\frac{3}{2}^-$~~   &~~0~~ &~~$\frac{3}{2}$~~ & Doublet & 1\\
& $c\bar{c}qq'q''$ &$|\mathbf{8}_{c\bar{c}},\bar{\mathbf{3}}_{qq'},\mathbf{3}_{q''}\rangle$& ${4.559^*}^{+0.084}_{-0.085}$~~  &~~$\frac{3}{2}^-$~~   &~~0~~ &~~$\frac{3}{2}$~~ & Doublet & 2\\\hline
& $c\bar{c}qq'q''$ &$|\mathbf{1}_{c\bar{c}},\bar{\mathbf{3}}_{qq'},\mathbf{3}_{q''}\rangle$& $4.383\pm0.068$~~  &~~$\frac{1}{2}^-$~~   &~~1~~ &~~$\frac{1}{2}$~~ & Sextet & 1\\
& $c\bar{c}qq'q''$ &$|\mathbf{1}_{c\bar{c}},\bar{\mathbf{3}}_{qq'},\mathbf{3}_{q''}\rangle$& ${4.673^*}^{+0.097}_{-0.098}$~~  &~~$\frac{3}{2}^-$~~   &~~1~~ &~~$\frac{1}{2}$~~ & Sextet & 2\\
& $c\bar{c}qq'q''$ &$|\mathbf{8}_{c\bar{c}},\mathbf{6}_{qq'},\mathbf{3}_{q''}\rangle$& ${4.576^*}^{+0.086}_{-0.087}$~~  &~~$\frac{1}{2}^-$~~   &~~1~~ &~~$\frac{1}{2}$~~ & Sextet & 3\\
& $c\bar{c}qq'q''$ &$|\mathbf{8}_{c\bar{c}},\mathbf{6}_{qq'},\mathbf{3}_{q''}\rangle$& $4.357\pm0.065$~~  &~~$\frac{3}{2}^-$~~   &~~1~~ &~~$\frac{1}{2}$~~ & Sextet & 4\\
& $c\bar{c}qq'q''$ &$|\mathbf{8}_{c\bar{c}},\bar{\mathbf{3}}_{qq'},\mathbf{3}_{q''}\rangle$& $4.470^{+0.076}_{-0.075}$~~  &~~$\frac{1}{2}^-$~~   &~~1~~ &~~$\frac{1}{2}$~~ & Sextet & 5\\
& $c\bar{c}qq'q''$ &$|\mathbf{8}_{c\bar{c}},\bar{\mathbf{3}}_{qq'},\mathbf{3}_{q''}\rangle$& ${4.589^*}^{+0.088}_{-0.087}$~~  &~~$\frac{3}{2}^-$~~   &~~1~~ &~~$\frac{1}{2}$~~ & Sextet & 6\\\hline
& $c\bar{c}qq'q''$ &$|\mathbf{1}_{c\bar{c}},\bar{\mathbf{3}}_{qq'},\mathbf{3}_{q''}\rangle$& ${4.673^*}^{+0.097}_{-0.098}$~~  &~~$\frac{1}{2}^-$~~   &~~1~~ &~~$\frac{3}{2}$~~ & Sextet & 1\\
& $c\bar{c}qq'q''$ &$|\mathbf{1}_{c\bar{c}},\bar{\mathbf{3}}_{qq'},\mathbf{3}_{q''}\rangle$& ${4.673^*}^{+0.097}_{-0.098}$~~  &~~$\frac{3}{2}^-$~~   &~~1~~ &~~$\frac{3}{2}$~~ & Sextet & 2\\
& $c\bar{c}qq'q''$ &$|\mathbf{1}_{c\bar{c}},\bar{\mathbf{3}}_{qq'},\mathbf{3}_{q''}\rangle$& ${4.673^*}^{+0.097}_{-0.098}$~~  &~~$\frac{5}{2}^-$~~   &~~1~~ &~~$\frac{3}{2}$~~ & Sextet & 3\\
& $c\bar{c}qq'q''$ &$|\mathbf{8}_{c\bar{c}},\bar{\mathbf{3}}_{qq'},\mathbf{3}_{q''}\rangle$& ${4.533^*}\pm0.082$~~  &~~$\frac{1}{2}^-$~~   &~~1~~ &~~$\frac{3}{2}$~~ & Sextet & 4\\
& $c\bar{c}qq'q''$ &$|\mathbf{8}_{c\bar{c}},\bar{\mathbf{3}}_{qq'},\mathbf{3}_{q''}\rangle$& ${4.565^*}\pm0.085$~~  &~~$\frac{3}{2}^-$~~   &~~1~~ &~~$\frac{3}{2}$~~ & Sextet & 5\\
& $c\bar{c}qq'q''$ &$|\mathbf{8}_{c\bar{c}},\bar{\mathbf{3}}_{qq'},\mathbf{3}_{q''}\rangle$& ${4.608^*}\pm0.089$~~  &~~$\frac{5}{2}^-$~~   &~~1~~ &~~$\frac{3}{2}$~~ & Sextet & 6\\\hline
& $ccqq'\bar{q''}$ &$|\mathbf{6}_{cc},\bar{\mathbf{3}}_{qq'},\bar{\mathbf{3}}_{\bar{q}''}\rangle$& ${4.642^*}\pm0.094$~~  &~~$\frac{1}{2}^-$~~   &~~0~~ &~~$\frac{1}{2}$~~ & Singlet & 1\\\hline
& $ccqq'\bar{q''}$ &$|\mathbf{6}_{cc},\bar{\mathbf{3}}_{qq'},\bar{\mathbf{3}}_{\bar{q}''}\rangle$& $4.523^{+0.082}_{-0.083}$~~  &~~$\frac{3}{2}^-$~~   &~~0~~ &~~$\frac{3}{2}$~~ & Singlet & 1\\\hline
& $ccqq'\bar{q''}$ &$|\bar{\mathbf{3}}_{cc},\mathbf{6}_{qq'},\bar{\mathbf{3}}_{\bar{q}''}\rangle$& ${4.592^*}^{+0.090}_{-0.089}$~~  &~~$\frac{1}{2}^-$~~   &~~1~~ &~~$\frac{1}{2}$~~ & Quartet & 1\\
& $ccqq'\bar{q''}$ &$|\bar{\mathbf{3}}_{cc},\mathbf{6}_{qq'},\bar{\mathbf{3}}_{\bar{q}''}\rangle$& ${4.567^*}^{+0.087}_{-0.088}$~~  &~~$\frac{3}{2}^-$~~   &~~1~~ &~~$\frac{1}{2}$~~ & Quartet & 2\\
& $ccqq'\bar{q''}$ &$|\bar{\mathbf{3}}_{cc},\bar{\mathbf{3}}_{qq'},\bar{\mathbf{3}}_{\bar{q}''}\rangle$& $4.451^{+0.075}_{-0.076}$~~  &~~$\frac{1}{2}^-$~~   &~~1~~ &~~$\frac{1}{2}$~~ & Quartet & 3\\
& $ccqq'\bar{q''}$ &$|\bar{\mathbf{3}}_{cc},\bar{\mathbf{3}}_{qq'},\bar{\mathbf{3}}_{\bar{q}''}\rangle$& $4.414^{+0.072}_{-0.073}$~~  &~~$\frac{3}{2}^-$~~   &~~1~~ &~~$\frac{1}{2}$~~ & Quartet & 4\\\hline
& $ccqq'\bar{q''}$ &$|\bar{\mathbf{3}}_{cc},\mathbf{6}_{qq'},\bar{\mathbf{3}}_{\bar{q}''}\rangle$& ${4.549^*}^{+0.086}_{-0.085}$~~  &~~$\frac{1}{2}^-$~~   &~~1~~ &~~$\frac{3}{2}$~~ & Triplet & 1\\
& $ccqq'\bar{q''}$ &$|\bar{\mathbf{3}}_{cc},\mathbf{6}_{qq'},\bar{\mathbf{3}}_{\bar{q}''}\rangle$& ${4.596^*}\pm0.090$~~  &~~$\frac{3}{2}^-$~~   &~~1~~ &~~$\frac{3}{2}$~~ & Triplet & 2\\
& $ccqq'\bar{q''}$ &$|\bar{\mathbf{3}}_{cc},\bar{\mathbf{3}}_{qq'},\bar{\mathbf{3}}_{\bar{q}''}\rangle$& ${4.656^*}^{+0.096}_{-0.097}$~~  &~~$\frac{5}{2}^-$~~   &~~1~~ &~~$\frac{3}{2}$~~ & Triplet & 3\\
\hline
           \bottomrule[1pt]\bottomrule[1pt]

\end{tabular}
\end{table*}
%\begin{sidewaystable}
%\centering
\begin{table*}[ht]
\renewcommand\arraystretch{1.5}
\caption{\label{spectra.p2} Hidden- and double-bottom pentaquark spectra (in GeV) in the classification of the heavy quark spin symmetry. }
\begin{tabular}{lc|c|c|c|c|c|c|c}
\toprule[1pt] \toprule[1pt]
      &Constituents~~   & Color structure &Mass~~    &~~$J^P$~~   &~~$s_{b\bar{b}}$ or $s_{bb}$~~ &~~$s_\ell$~~& Multiplet or singlet &~~Label~~ \\ \hline
& $b\bar{b}qq'q''$ &$|\mathbf{1}_{b\bar{b}},\bar{\mathbf{3}}_{qq'},\mathbf{3}_{q''}\rangle$& $10.98\pm0.06$~~  &~~$\frac{1}{2}^-$~~   &~~0~~ &~~$\frac{1}{2}$~~ & Triplet & 1\\
& $b\bar{b}qq'q''$ &$|\mathbf{8}_{b\bar{b}},\mathbf{6}_{qq'},\mathbf{3}_{q''}\rangle$& ${11.26^*}\pm0.09$~~  &~~$\frac{1}{2}^-$~~   &~~0~~ &~~$\frac{1}{2}$~~ & Triplet & 2\\
& $b\bar{b}qq'q''$ &$|\mathbf{8}_{b\bar{b}},\bar{\mathbf{3}}_{qq'},\mathbf{3}_{q''}\rangle$& $11.15\pm0.08$~~  &~~$\frac{1}{2}^-$~~   &~~0~~ &~~$\frac{1}{2}$~~ & Triplet & 3\\\hline
& $b\bar{b}qq'q''$ &$|\mathbf{1}_{b\bar{b}},\bar{\mathbf{3}}_{qq'},\mathbf{3}_{q''}\rangle$& ${11.27^*}\pm0.09$~~  &~~$\frac{3}{2}^-$~~   &~~0~~ &~~$\frac{3}{2}$~~ & Doublet & 1\\
& $b\bar{b}qq'q''$ &$|\mathbf{8}_{b\bar{b}},\bar{\mathbf{3}}_{qq'},\mathbf{3}_{q''}\rangle$& ${11.24^*}\pm0.09$~~  &~~$\frac{3}{2}^-$~~   &~~0~~ &~~$\frac{3}{2}$~~ & Doublet & 2\\\hline
& $b\bar{b}qq'q''$ &$|\mathbf{1}_{b\bar{b}},\bar{\mathbf{3}}_{qq'},\mathbf{3}_{q''}\rangle$& $11.04\pm0.07$~~  &~~$\frac{1}{2}^-$~~   &~~1~~ &~~$\frac{1}{2}$~~ & Sextet & 1\\
& $b\bar{b}qq'q''$ &$|\mathbf{1}_{b\bar{b}},\bar{\mathbf{3}}_{qq'},\mathbf{3}_{q''}\rangle$& ${11.34^*}\pm0.10$~~  &~~$\frac{3}{2}^-$~~   &~~1~~ &~~$\frac{1}{2}$~~ & Sextet & 2\\
& $b\bar{b}qq'q''$ &$|\mathbf{8}_{b\bar{b}},\mathbf{6}_{qq'},\mathbf{3}_{q''}\rangle$& ${11.25^*}\pm0.09$~~  &~~$\frac{1}{2}^-$~~   &~~1~~ &~~$\frac{1}{2}$~~ & Sextet & 3\\
& $b\bar{b}qq'q''$ &$|\mathbf{8}_{b\bar{b}},\mathbf{6}_{qq'},\mathbf{3}_{q''}\rangle$& $11.03\pm0.07$~~  &~~$\frac{3}{2}^-$~~   &~~1~~ &~~$\frac{1}{2}$~~ & Sextet & 4\\
& $b\bar{b}qq'q''$ &$|\mathbf{8}_{b\bar{b}},\bar{\mathbf{3}}_{qq'},\mathbf{3}_{q''}\rangle$& $11.15\pm0.04$~~  &~~$\frac{1}{2}^-$~~   &~~1~~ &~~$\frac{1}{2}$~~ & Sextet & 5\\
& $b\bar{b}qq'q''$ &$|\mathbf{8}_{b\bar{b}},\bar{\mathbf{3}}_{qq'},\mathbf{3}_{q''}\rangle$& ${11.26^*}\pm0.05$~~  &~~$\frac{3}{2}^-$~~   &~~1~~ &~~$\frac{1}{2}$~~ & Sextet & 6\\\hline
& $b\bar{b}qq'q''$ &$|\mathbf{1}_{b\bar{b}},\bar{\mathbf{3}}_{qq'},\mathbf{3}_{q''}\rangle$& ${11.34^*}\pm0.10$~~  &~~$\frac{1}{2}^-$~~   &~~1~~ &~~$\frac{3}{2}$~~ & Sextet & 1\\
& $b\bar{b}qq'q''$ &$|\mathbf{1}_{b\bar{b}},\bar{\mathbf{3}}_{qq'},\mathbf{3}_{q''}\rangle$& ${11.34^*}\pm0.10$~~  &~~$\frac{3}{2}^-$~~   &~~1~~ &~~$\frac{3}{2}$~~ & Sextet & 2\\
& $b\bar{b}qq'q''$ &$|\mathbf{1}_{b\bar{b}},\bar{\mathbf{3}}_{qq'},\mathbf{3}_{q''}\rangle$& ${11.34^*}\pm0.10$~~  &~~$\frac{5}{2}^-$~~   &~~1~~ &~~$\frac{3}{2}$~~ & Sextet & 3\\
& $b\bar{b}qq'q''$ &$|\mathbf{8}_{b\bar{b}},\bar{\mathbf{3}}_{qq'},\mathbf{3}_{q''}\rangle$& ${11.24^*}\pm0.08$~~  &~~$\frac{1}{2}^-$~~   &~~1~~ &~~$\frac{3}{2}$~~ & Sextet & 4\\
& $b\bar{b}qq'q''$ &$|\mathbf{8}_{b\bar{b}},\bar{\mathbf{3}}_{qq'},\mathbf{3}_{q''}\rangle$& ${11.25^*}\pm0.09$~~  &~~$\frac{3}{2}^-$~~   &~~1~~ &~~$\frac{3}{2}$~~ & Sextet & 5\\
& $b\bar{b}qq'q''$ &$|\mathbf{8}_{b\bar{b}},\bar{\mathbf{3}}_{qq'},\mathbf{3}_{q''}\rangle$& ${11.27^*}\pm0.09$~~  &~~$\frac{5}{2}^-$~~   &~~1~~ &~~$\frac{3}{2}$~~ & Sextet & 6\\\hline
& $bbqq'\bar{q''}$ &$|\mathbf{6}_{bb},\bar{\mathbf{3}}_{qq'},\bar{\mathbf{3}}_{\bar{q}''}\rangle$& ${11.30^*}^{+0.09}_{-0.10}$~~  &~~$\frac{1}{2}^-$~~   &~~0~~ &~~$\frac{1}{2}$~~ & Singlet & 1\\\hline
& $bbqq'\bar{q''}$ &$|\mathbf{6}_{bb},\bar{\mathbf{3}}_{qq'},\bar{\mathbf{3}}_{\bar{q}''}\rangle$& $11.18\pm0.08$~~  &~~$\frac{3}{2}^-$~~   &~~0~~ &~~$\frac{3}{2}$~~ & Singlet & 1\\\hline
& $bbqq'\bar{q''}$ &$|\bar{\mathbf{3}}_{bb},\mathbf{6}_{qq'},\bar{\mathbf{3}}_{\bar{q}''}\rangle$& ${11.25^*}^{+0.08}_{-0.09}$~~  &~~$\frac{1}{2}^-$~~   &~~1~~ &~~$\frac{1}{2}$~~ & Quartet & 1\\
& $bbqq'\bar{q''}$ &$|\bar{\mathbf{3}}_{bb},\mathbf{6}_{qq'},\bar{\mathbf{3}}_{\bar{q}''}\rangle$& ${11.24^*}^{+0.08}_{-0.09}$~~  &~~$\frac{3}{2}^-$~~   &~~1~~ &~~$\frac{1}{2}$~~ & Quartet & 2\\
& $bbqq'\bar{q''}$ &$|\bar{\mathbf{3}}_{bb},\bar{\mathbf{3}}_{qq'},\bar{\mathbf{3}}_{\bar{q}''}\rangle$& $11.12^{+0.08}_{-0.07}$~~  &~~$\frac{1}{2}^-$~~   &~~1~~ &~~$\frac{1}{2}$~~ & Quartet & 3\\
& $bbqq'\bar{q''}$ &$|\bar{\mathbf{3}}_{bb},\bar{\mathbf{3}}_{qq'},\bar{\mathbf{3}}_{\bar{q}''}\rangle$& $11.09^{+0.07}_{-0.08}$~~  &~~$\frac{3}{2}^-$~~   &~~1~~ &~~$\frac{1}{2}$~~ & Quartet & 4\\\hline
& $bbqq'\bar{q''}$ &$|\bar{\mathbf{3}}_{bb},\mathbf{6}_{qq'},\bar{\mathbf{3}}_{\bar{q}''}\rangle$& ${11.21^*}^{+0.09}_{-0.08}$~~  &~~$\frac{1}{2}^-$~~   &~~1~~ &~~$\frac{3}{2}$~~ & Triplet & 1\\
& $bbqq'\bar{q''}$ &$|\bar{\mathbf{3}}_{bb},\mathbf{6}_{qq'},\bar{\mathbf{3}}_{\bar{q}''}\rangle$& ${11.25^*}\pm0.09$~~  &~~$\frac{3}{2}^-$~~   &~~1~~ &~~$\frac{3}{2}$~~ & Triplet & 2\\
& $bbqq'\bar{q''}$ &$|\bar{\mathbf{3}}_{bb},\bar{\mathbf{3}}_{qq'},\bar{\mathbf{3}}_{\bar{q}''}\rangle$& ${11.30^*}\pm0.09$~~  &~~$\frac{5}{2}^-$~~   &~~1~~ &~~$\frac{3}{2}$~~ & Triplet & 3\\
\hline
           \bottomrule[1pt]\bottomrule[1pt]

\end{tabular}
\end{table*}

\begin{table*}[ht]
\renewcommand\arraystretch{1.5}
\caption{\label{tab:Ratio} The relative weights of various hidden- and double-charmed pentaquark decay processes under the heavy quark symmetry. The similar ratios for hidden- and double-bottom states can be readily obtained by the replacements of $c\to b,~P_c \to P_b,~P_{cc}\to P_{bb},~J/\psi\to\Upsilon,~\eta_c\to\eta_b$, {\rm and} $\Xi_{cc}\to\Xi_{bb}$. }

\begin{tabular}{|cc|c|c|c|}\hline\hline
\multicolumn{5}{|c|}{$P_c\to H_{[c\bar{c}]}+P$}
\\\hline
&$\Gamma_i/\Gamma_j$ & $R$  & $\Gamma_i/\Gamma_j$ & $R$
\\\hline
&$\frac{\Gamma(P_c(J^P=\frac{1}{2}^-,s_{c\bar{c}}=1,s_\ell=\frac{3}{2})\to (J/\psi p)|_{J'=\frac{3}{2}})}{\Gamma(P_c(J^P=\frac{1}{2}^-,s_{c\bar{c}}=1,s_\ell=\frac{3}{2})\to (J/\psi p)|_{J'=\frac{1}{2}})}$ &
$\frac{1}{8}$ &$\frac{\Gamma(P_c(J^P=\frac{3}{2}^-,s_{c\bar{c}}=1,s_\ell=\frac{3}{2})\to (J/\psi p)|_{J'=\frac{3}{2}})}{\Gamma(P_c(J^P=\frac{3}{2}^-,s_{c\bar{c}}=1,s_\ell=\frac{3}{2})\to (J/\psi p)|_{J'=\frac{1}{2}})}$ &
$\frac{4}{5}$
\\\hline
&$\frac{\Gamma(P_c(J^P=\frac{3}{2}^-,s_{c\bar{c}}=1,s_\ell=\frac{3}{2})\to (J/\psi p)|_{J'=\frac{1}{2}})}{\Gamma(P_c(J^P=\frac{1}{2}^-,s_{c\bar{c}}=1,s_\ell=\frac{3}{2})\to (J/\psi p)|_{J'=\frac{1}{2}})}$ &
$\frac{5}{8}$ &$\frac{\Gamma(P_c(J^P=\frac{3}{2}^-,s_{c\bar{c}}=1,s_\ell=\frac{3}{2})\to (J/\psi \Delta)|_{J'=\frac{3}{2}})}{\Gamma(P_c(J^P=\frac{1}{2}^-,s_{c\bar{c}}=1,s_\ell=\frac{3}{2})\to (J/\psi p)|_{J'=\frac{1}{2}})}$ &
$\frac{121}{200}$
\\\hline
&$\frac{\Gamma(P_c(J^P=\frac{3}{2}^-,s_{c\bar{c}}=1,s_\ell=\frac{3}{2})\to (J/\psi \Delta)|_{J'=\frac{5}{2}})}{\Gamma(P_c(J^P=\frac{1}{2}^-,s_{c\bar{c}}=1,s_\ell=\frac{3}{2})\to (J/\psi p)|_{J'=\frac{1}{2}})}$&$\frac{27}{100}$ &
$\frac{\Gamma(P_c(J^P=\frac{3}{2}^-,s_{c\bar{c}}=1,s_\ell=\frac{3}{2})\to (J/\psi \Delta)|_{J'=\frac{1}{2}})}{\Gamma(P_c(J^P=\frac{1}{2}^-,s_{c\bar{c}}=1,s_\ell=\frac{3}{2})\to (J/\psi p)|_{J'=\frac{1}{2}})}$ &
$\frac{1}{4}$
\\\hline
&$\frac{\Gamma(P_c(J^P=\frac{1}{2}^-,s_{c\bar{c}}=1,s_\ell=\frac{3}{2})\to (J/\psi \Delta)|_{J'=\frac{1}{2}})}{\Gamma(P_c(J^P=\frac{1}{2}^-,s_{c\bar{c}}=1,s_\ell=\frac{3}{2})\to (J/\psi p)|_{J'=\frac{1}{2}})}$&$\frac{5}{8}$ &
$\frac{\Gamma(P_c(J^P=\frac{1}{2}^-,s_{c\bar{c}}=1,s_\ell=\frac{3}{2})\to (J/\psi \Delta)|_{J'=\frac{3}{2}})}{\Gamma(P_c(J^P=\frac{1}{2}^-,s_{c\bar{c}}=1,s_\ell=\frac{3}{2})\to (J/\psi p)|_{J'=\frac{1}{2}})}$ &
$\frac{1}{2}$
\\\hline
&$\frac{\Gamma(P_c(J^P=\frac{1}{2}^-,s_{c\bar{c}}=0,s_\ell=\frac{1}{2})\to \eta_c \Delta)}{\Gamma(P_c(J^P=\frac{1}{2}^-,s_{c\bar{c}}=0,s_\ell=\frac{1}{2})\to \eta_c  p)}$&$1$ &
$\frac{\Gamma(P_c(J^P=\frac{3}{2}^-,s_{c\bar{c}}=0,s_\ell=\frac{3}{2})\to \eta_c \Delta)}{\Gamma(P_c(J^P=\frac{3}{2}^-,s_{c\bar{c}}=0,s_\ell=\frac{3}{2})\to \eta_c  p)}$ &
$1$
\\\hline
\multicolumn{5}{|c|}{$P_{cc}\to H_{cc}+P$}
\\\hline
&$\Gamma_i/\Gamma_j$ & $R$  & $\Gamma_i/\Gamma_j$ & $R$
\\\hline
&$\frac{\Gamma(P_{cc}(J^P=\frac{1}{2}^-,s_{cc}=0,s_\ell=\frac{1}{2})\to \Xi_{cc}(J^P=\frac{3}{2}^-)\pi}{\Gamma(P_{cc}(J^P=\frac{1}{2}^-,s_{cc}=0,s_\ell=\frac{1}{2})\to \Xi_{cc}(J^P=\frac{1}{2}^-)\pi}$ &
$1$ &$\frac{\Gamma(P_{cc}(J^P=\frac{1}{2}^-,s_{cc}=1,s_\ell=\frac{1}{2})\to \Xi_{cc}(J^P=\frac{1}{2}^-)\pi}{\Gamma(P_{cc}(J^P=\frac{3}{2}^-,s_{cc}=1,s_\ell=\frac{1}{2})\to \Xi_{cc}(J^P=\frac{3}{2}^-)\pi}$ &
$1$\\\hline\hline
\end{tabular}
\end{table*}

Note, in our approach, $\beta$ and $R$ are free parameters. The linear confinement potential is also unknown because of the lack of the information of $b_1$ and $b_0$. For simplicity, in the calculation we do not consider the linear confinement potential contribution but let parameters $\beta$ and $R$ vary. The color and spin matrix elements are given in Tabs.~\ref{color.1}, \ref{color.2}, \ref{spin10.5.1}, and \ref{spin9.5.1}.
The constituent quark masses are chosen as employed in \cite{Maiani:2004vq,Ali:2011ug,Wang:2016tsi,Wang:2017vnc}, i.e. $m_{q}=305\pm20$ MeV,  $m_{c}=1670\pm10$ MeV, and $m_{b}=5008\pm10$ MeV for mesons (Set I); $m_{q}=362\pm20$ MeV,  $m_{c}=1721\pm10$ MeV, and $m_{b}=5050\pm10$ MeV for baryons (Set II). Therein we vary the light quark mass by 20 MeV while the heavy quark mass by 10 MeV. Thus the central values of constituent quarks masses threshold become 3.950 GeV for  charm tetraquarks, 10.63 GeV for bottom tetraquarks, 4.528 GeV for  charm pentaquarks, and 11.19 GeV for bottom pentaquarks. The couplings are chosen as $C^{qq}/m_q^2=193\pm19$ MeV, $C^{cq}/(m_q m_c)=23\pm2$ MeV, $C^{bq}/(m_q m_b)=12\pm1$ MeV and $C^{q\bar{q}}/m_q^2=318\pm32$ MeV, $C^{c\bar{q}}/(m_q m_c)=69\pm7$ MeV, $C^{b\bar{q}}/(m_q m_b)=23\pm2$ MeV, $C^{c\bar{c}}/m_c^2=57\pm6$ MeV, $C^{b\bar{b}}/m_b^2=31\pm3$ MeV~\cite{Xing:2018bqt}, where 10 percent uncertainty is implied.

Through the calculation, we find that the optimal value of $\beta$ is around 50 MeV, while the optimal value of $R$ is around $1$ fm for doubly heavy flavor tetraquarks. For hidden heavy flavor tetraquark, hidden heavy flavor pentaquark, and doubly heavy flavor pentaquark, the optimal value of $\beta$ is around 100 MeV, while the optimal value of $R$ is 0 which leads to divergence. We find the $Q\bar{Q} q q' q''$ and $Q\bar{Q} q \bar{q}'$ system becomes more attractive than the $QQ q q'\bar{q}'' $ and $\bar{Q}\bar{Q} q q'$ systems when the heavy flavor distance $R$ tends to small. To avoid the divergence, we set a typical value of 1 fm for $R$ for hidden heavy flavor tetraquarks and heavy flavor pentaquarks. Noticing the test function in the variational method may induce some uncertainties, we estimate this type of error induced by calculating the variance, the square of the standard deviation $\sigma=\sqrt{\langle \hat{{\mathcal H}}^2\rangle-\langle \hat{{\mathcal H}}\rangle^2}$, which is about 100 MeV for hidden heavy flavor tetraquarks, hidden heavy flavor pentaquarks and doubly heavy flavor pentaquarks, while about 50 MeV for doubly heavy flavor tetraquarks.

The spectra of the  pentaquarks are given in Tabs.~\ref{spectra.p} and \ref{spectra.p2}, where the values of the parameters are chosen as Set II. These pentaquarks are grouped into different multiplets or singlets under the heavy quark symmetry. Here we only focus on the S-wave states and ignore the orbitally excited states. Considering of the LHCb data for the hidden charm pentaquarks, the $P_c(4312)^+$ may be assigned to the ground state with spin-parity $\frac{1}{2}^-$ or $\frac{3}{2}^-$; the $P_c(4440)^+$ and $P_c(4457)^+$
may be assigned to the excited states with $\frac{1}{2}^-$ or $\frac{3}{2}^-$. Besides, the wide resonance $P_c(4380)^+$ may be assigned to the excited state with $\frac{1}{2}^-$, or the ground state with $\frac{3}{2}^-$, or from the interference by two  states with $\frac{1}{2}^-$. Of course, one may notice that the $P_c(4440)^+$  and $P_c(4457)^+$ state might also be the ground state of $\frac{5}{2}^-$ or the excited states of $\frac{3}{2}^-$ if we employ the parameter values of Set I.

The spectra of the  hidden charm tetraquarks  are (in  GeV)
\begin{align}\label{mass1}
 m(c\bar{c}qq')&= \left\{ \begin{array} {ll}
  3.860\pm0.051,  & J^P=0^+ , {s_{c\bar{c}}}=0, {s_\ell}=0, \\
3.735\pm0.039,  & J^P=1^+ , {s_{c\bar{c}}}=0, {s_\ell}=1, \\
3.956\pm0.061,  & J^P=1^+ , {s_{c\bar{c}}}=1, {s_\ell}=0, \\
 4.023^{+0.068}_{-0.067},  & J^P=0^+ , {s_{c\bar{c}}}=1, {s_\ell}=1, \\
3.854^{+0.051}_{-0.050},  & J^P=1^+ , {s_{c\bar{c}}}=1, {s_\ell}=1, \\
4.096\pm0.075,  & J^P=2^+ , {s_{c\bar{c}}}=1, {s_\ell}=1 .\end{array} \right.
 \end{align}

Considering of the available data for the hidden charm tetraquarks, the $X(3823)$ state~\cite{Bhardwaj:2013rmw} may be thought of the ground  hidden charm tetraquark state with $0^+$; $Z_c(3900)$~\cite{Ablikim:2013mio} may be assigned as the  hidden charm tetraquark state with $1^+$; and $Z_c(4020)$~\cite{Ablikim:2013wzq} may be assigned as one of the excited states of hidden charm tetraquark state with $0^+$ or $1^+$, or the ground hidden charm tetraquark state with $2^+$ .

 The spectra of the doubly charm tetraquarks  become (in  GeV)
\begin{align}\label{mass2}
 m(cc\bar{q}\bar{q'})&= \left\{ \begin{array} {ll}
3.986^{+0.069}_{-0.068},  & J^P=0^+ , {s_{cc}}=0, {s_\ell}=0, \\
3.923\pm0.059,  & J^P=0^+ , {s_{cc}}=1, {s_\ell}=1, \\
3.957^{+0.063}_{-0.062},  & J^P=1^+ , {s_{cc}}=1, {s_\ell}=1, \\
4.026\pm0.069,  & J^P=2^+ , {s_{cc}}=1, {s_\ell}=1 .\end{array} \right.
 \end{align}
From the above analysis, there exist three separated singlets and one triplet for hidden charm
tetraquarks, while exist one singlet and one triplet for doubly charmed tetraquarks.

Similarly, the spectra of the  hidden bottom tetraquarks  become (in  GeV)
\begin{align}\label{mass1}
 m(b\bar{b}qq')&=\left\{ \begin{array} {ll}
  10.66\pm0.05,  & J^P=0^+ , {s_{b\bar{b}}}=0, {s_\ell}=0, \\
10.70\pm0.05,  & J^P=1^+ , {s_{b\bar{b}}}=0, {s_\ell}=1, \\
10.77\pm0.06,  & J^P=1^+ , {s_{b\bar{b}}}=1, {s_\ell}=0, \\
 10.83\pm0.07,  & J^P=0^+ , {s_{b\bar{b}}}=1, {s_\ell}=1, \\
10.74\pm0.06,  & J^P=1^+ , {s_{b\bar{b}}}=1, {s_\ell}=1, \\
10.82\pm0.07,  & J^P=2^+ , {s_{b\bar{b}}}=1, {s_\ell}=1 .\end{array} \right.
 \end{align}

The spectra of the doubly bottom tetraquarks  are (in  GeV)
\begin{align}\label{mass1}
 m(bb\bar{q}\bar{q'})&= \left\{ \begin{array} {ll}
 10.68\pm0.07,  & J^P=0^+ , {s_{bb}}=0, {s_\ell}=0, \\
10.65\pm0.06,  & J^P=0^+ , {s_{bb}}=1, {s_\ell}=1, \\
10.66\pm0.06,  & J^P=1^+ , {s_{bb}}=1, {s_\ell}=1, \\
10.68\pm0.07,  & J^P=2^+ , {s_{bb}}=1, {s_\ell}=1 .\end{array} \right.
 \end{align}

We calculate the spectra of the S-wave multi-quark  states with two heavy flavors. The orbitally excited states are not considered here.  To pin down these multi-quark states, finding some experimentally accessible channels are necessary and important. For the hidden heavy flavor pentaquarks, based on certain analysis, we believe $\Lambda_b\to J/\psi+p^+ (\Delta^+) +K^-$, $\Lambda_b\to J/\psi+n +\bar{K}^0$, $pp(\bar{p}) \to \Upsilon+p^+ (\Delta^+)+X$ and $pp(\bar{p}) \to \Lambda_{c/b}+\bar{D}(\bar{B})+X$ processes are hopefully detectable in currently running experiments. For the doubly heavy flavor pentaquarks, the $pp(\bar{p}) \to \Lambda_{c/b}+D(B)+X$ process might be accessible, and
for the doubly heavy flavor tetraquarks, one may pay attention to $pp(\bar{p}) \to \bar{\Lambda}+\Xi_c+X$ and $pp(\bar{p}) \to \bar{p}+\Xi_b+X$ processes.

Note, one may get the information of relative ratios of different processes via the analysis of heavy quark spin symmetry \cite{Isgur:1991wq,Sakai:2019qph}. Take the two-body exclusive decay of hidden charm pentaquark to S-wave a charmonium and a light baryon as an example, the decay widths in the heavy quark symmetry tells
\begin{eqnarray}
\Gamma&\propto&(2s_\ell+1)(2J'+1)\left|\left\{\begin{array}{ccc}
                                        L & s'_\ell & s_\ell \\
                                        s_{c\bar{c}} & J & J'
                                      \end{array}\right\}\right|^2,
\label{eq:decay2}
\end{eqnarray}
in terms of the 6j symbols of Clebsch-Gordan coefficients. Here, $L$ denotes the orbital angular momentum of the emitted light baryon; $s'_\ell$ is the light degree of freedom in final states; $J$ represents the total angular momentum of the pentaquark while $J'$ is the total angular momentum of the charmonium and light baryon.

To summarize, we give some explanation on the calculation yields in the following:
\begin{itemize}

\item Considering the LHCb measurements of $P_c(X)\to J/\psi+p$, the spin of heavy quark pair $s_{c\bar{c}}$ can only be 1 under the heavy quark symmetry. Possible choices for $P_c(4312)$, $P_c(4440)$ and $P_c(4457)$ imply that they all belong to the sextet with $s_{c\bar{c}}=1$ and $s_\ell=\frac{3}{2}$ in Tab.~\ref{spectra.p}. So their spin-parity $J^P$ cold be either $\frac{1}{2}^-$ or $\frac{3}{2}^-$. Due to the parity conservation, the orbital angular momentum of the light baryon can only be odd. If one ignores the phase space effect, the ratios of different decay channels under the heavy quark symmetry can be obtained, which are listed in Tab.~\ref{tab:Ratio}.

\item For double-charm pentaquark decays, we present the relative ratios in Tab.~\ref{tab:Ratio}.
For tetraquarks decays, the $T_{c}(J^P=1^+,s_{cc}=1,s_\ell=1)\to J/\psi\pi$ processes are legitimate under the heavy quark symmetry and thus the $Z_c$ states may be assigned to tetraquarks $T_{c}(J^P=1^+,s_{cc}=1,s_\ell=1)$. Moreover, there are still many of other channels are  allowed in heavy quark symmetry, such as $T_{c}(J^P=1^+,s_{cc}=1,s_\ell=1)\to J/\psi\rho$, $T_{c}(J^P=0^+,s_{cc}=0,s_\ell=0)\to \eta_c\rho$, $T_{c}(J^P=2^+,s_{cc}=1,s_\ell=1)\to J/\psi\rho$, which may be explored in experiment for the study of exotic states.

\item Different theoretical frameworks may lead to different conclusions on the LHCb $P_c$ states. The molecular pentaquark model is one of the attractive options. Based on it, as an example, the molecular states $\Sigma_c D$ with $J^P=\frac{1}{2}^-$, $\Sigma_c D^*$ with $J^P=\frac{3}{2}^-$ and $\Sigma_c D^*$ with $J^P=\frac{1}{2}^-$ can well fit to the data \cite{Huang:2019jlf,Huang:2018wed}.

\end{itemize}

\section{Conclusion}

In this work we calculated the spectra of the hidden heavy flavor and doubly heavy flavor pentaquarks, hidden heavy flavor and doubly heavy flavor tetraquarks by virtue of the variational method.  We adopted the model for multiquark system similar to a hydrogen molecule but with SU(3) color interactions. According to our results, the Set II, $P_c(4312)^+$ state observed by LHCb Collaboration could be a ground state of the multiquark system with spin-parity $\frac{1}{2}^-$ or $\frac{3}{2}^-$, while the $P_c(4440)^+$ and $P_c(4457)^+$ might be excited states with
$\frac{1}{2}^-$. These three pentaquarks may  all belong to the sextet with $s_{c\bar{c}}=1$ and $s_\ell=\frac{3}{2}$. The hydrogen-like model indicates that the $Q\bar{Q} q q' q''$ and $Q\bar{Q} q \bar{q}'$ systems become more attractive and stable than the $QQ q q'\bar{q}'' $ and $\bar{Q}\bar{Q} q q'$ systems when the heavy flavor distance $R$ shrinks. We presented some promising decay channels of those multiquark states considered, which are left for experiment confirmation. A deeper and wider investigation on multiquark system shall no doubt enlighten us on the exotic hadrons and the nature of QCD.

\section*{Acknowledgments}

 The authors thank the useful discussions with Prof. Jialun Ping. This work was
supported in part by the National Natural Science Foundation of
China under Grant No.~11705092, 11635009, and 11675080,  by Natural Science Foundation of Jiangsu under
Grant No.~BK20171471, by the Ministry of Science and Technology of the Peoples' Republic of China(2015CB856703).

\end{document}